\documentclass[preprint,11pt]{elsarticle}% Math and Physical Sciences Numbered Reference Style 
\usepackage[top=0.7in, right=1in, left=1in, bottom=1in]{geometry}
\usepackage{graphicx}%
\usepackage{multirow}%
\usepackage{amsmath,amssymb,amsfonts,amsthm}%
\usepackage{amsthm}%
\usepackage{mathrsfs}%
\usepackage[title]{appendix}%
\usepackage[dvipsnames]{xcolor}%
\usepackage{caption,subcaption,float}
\usepackage{textcomp}%
\usepackage{hyperref,url}
\usepackage{manyfoot}%
\usepackage{booktabs}%
\usepackage{algorithm}%
\usepackage{listings}%
\usepackage{pgfplots}
\usepackage{tikz}
\usepackage{natbib}
\usepackage{algorithm}
\usepackage{algorithmic}
\usepackage{hypernat}
\hypersetup{
    colorlinks=true,
    linkcolor=blue,
    citecolor=blue,
    filecolor=magenta,      
    urlcolor=blue,
    pdftitle={DilectricPaper},
    pdfpagemode=FullScreen,
}

\usepackage{xpatch}
\xpatchcmd{\citet}{\@citep}{}{}{}

\definecolor{codegreen}{rgb}{0,0.6,0}
\definecolor{codegray}{rgb}{0.5,0.5,0.5}
\definecolor{codepurple}{rgb}{0.58,0,0.82}
\definecolor{backcolour}{rgb}{0.95,0.95,0.92}

\lstdefinestyle{mypythonstyle}{
    backgroundcolor=\color{backcolour},
    commentstyle=\color{codegreen},
    keywordstyle=\color{magenta},
    numberstyle=\tiny\color{codegray},
    stringstyle=\color{codepurple},
    basicstyle=\ttfamily\footnotesize,
    breakatwhitespace=false,
    breaklines=true,
    captionpos=b,
    keepspaces=true,
    numbers=left,
    numbersep=5pt,
    frame=lines,
    showspaces=false,
    showstringspaces=false,
    showtabs=false,
    tabsize=2,
    language=Python
}

\makeatletter
\def\ps@pprintTitle{%
 \let\@oddhead\@empty
 \let\@evenhead\@empty
 \def\@oddfoot{\reset@font\hfil\thepage\hfil}
 \let\@evenfoot\@oddfoot
}
\makeatother

\newtheoremstyle{remarkstyle}
  {3pt} % Space above
  {3pt} % Space below
  {\itshape} % Body font
  {} % Indent amount
  {\bfseries} % Theorem head font
  {:} % Punctuation after theorem head
  {.5em} % Space after theorem head
  {} % Theorem head spec

\theoremstyle{remarkstyle}
\newtheorem{remark}{Remark}%

\definecolor{lightblue}{RGB}{240,248,255}

\usepackage{listings}
\DeclareCaptionFormat{listing}{\raggedright#1#2#3}
\DeclareCaptionFormat{listing}{\raggedright#1#2#3}
\def\bfx{{\bf x}}
\def\bfy{{\bf y}}

\def\bfE{{\bf E}}

\def\bfC{{\bf C}}
\def\bfH{{\bf H}}
\def\bfI{{\bf I}}

\def\bfN{{\bf N}}

\def\bfS{{\bf S}}

\def\bfX{{\bf X}}
\def\bfF{{\bf F}}

\def\bfD{{\bf D}}

\def\bfe{{\bf e}}

\def\bfw{\mbox{\boldmath $\omega$}}

\def\Atan{\mbox{\boldmath$\mathcal{A}$}}

\def\Ktan{\mbox{\boldmath$\mathcal{K}$}}
\def\Jtan{\mbox{\boldmath$\mathcal{J}$}}

\def\Atan{\mbox{\boldmath$\mathcal{A}$}}

\def\e0{\varepsilon_0}

\def\obfF{\overline{\bfF}}

\def\s0{\sigma_0}

\def\obfF{\overline{\bfF}}

\long\def\symbolfootnote[#1]#2{\begingroup%
\def\thefootnote{\fnsymbol{footnote}}\footnote[#1]{#2}\endgroup}

\def\Atan{\mbox{\boldmath$\mathcal{A}$}}
\def\Btan{\mbox{\boldmath$\mathcal{B}$}}
\def\Ktan{\mbox{\boldmath$\mathcal{K}$}}
\def\Jtan{\mbox{\boldmath$\mathcal{J}$}}

\pgfplotsset{compat=1.16}
\begin{document}

\begin{frontmatter}

\title{Hybrid finite element implementation of two-potential constitutive modeling of dielectric elastomers}

\author[1]{Kamalendu Ghosh\corref{equal}}\ead{kamalendugho@micron.com}

\author[2]{Bhavesh Shrimali\corref{equal}\fnref{fn1}}\ead{bhavesh.shrimali@gmail.com}

\cortext[equal]{These authors contributed equally to this work.}
\address[1]{Equipment Development R\&D Engineer, Micron Technology, 8000 S Federal Way, Boise, ID 83716, USA}

\address[2]{Lead Scientist, Kimberly-Clark, 2100 Winchester Rd, Neenah, WI 54956, USA}
\fntext[fn1]{Work was carried out while at UIUC.}

\begin{abstract}
    Dielectric elastomers are increasingly studied for their potential in soft
robotics, actuators, and haptic devices. Under time-dependent loading, they
dissipate energy via viscous deformation and friction in electric polarization.
However, most constitutive models and finite element (FE) implementations
consider only mechanical dissipation because mechanical relaxation times are
much larger than electric ones. Accounting for electric dissipation is crucial
when dealing with alternating electric fields. \citet{ghosh2021two} proposed a
fully coupled three-dimensional constitutive model for isotropic,
incompressible dielectric elastomers. We critically investigate their numerical
scheme for solving the initial boundary value problem (IBVP) describing the
time-dependent behavior. We find that their fifth-order explicit Runge-Kutta
time discretization may require excessively small or unphysical time steps for
realistic simulations due to the stark contrast in mechanical and electric
relaxation times. To address this, we present a stable implicit
time-integration algorithm that overcomes these constraints. We implement this
algorithm with a conforming FE discretization to solve the IBVP and present the
mixed-FE formulation implemented in FEniCSx. We demonstrate that the scheme is
robust, accurate, and capable of handling finite deformations,
incompressibility, and general time-dependent loading. Finally, we validate our
code against experimental data for VHB 4910 under complex time-dependent
electromechanical loading, as studied by \citet{hossain2015comprehensive}.
\end{abstract}


\end{frontmatter}

% \\\\
% \textbf{Yet to finalize}

% \\\\

% A detailed step-by-step procedure of a FEniCSx and Abaqus UEL implementation of the mixed-FE formulation of the governing equations is spelled out. The scheme makes use of a conforming Crouzeix–Raviart FE discretization of space and a high-order accurate explicit Runge–Kutta discretization of time. The combination of these two types of discretizations results into a robust scheme that is capable of handling finite deformations, the incompressibility constraint of the rubber and general time-dependent loading conditions. In the last part, the Abaqus UEL and FEniCSx formulations are deployed to fully describe the electromechanical behavior of VHB 4910 (from 3M) under a complex time-dependent electromechanical load as experimentally studied in \cite{hossain2015comprehensive}.}

% \keywords{Dielectric Elastomer, Dissipative Solids, Internal Variables, Finite Element}

% \maketitle
\section{Introduction}\label{Sec:Introduction}
% \textbf{Why dielectric elastomers??\\}
Dielectric elastomers are soft materials that deform significantly when subjected to electric fields. These materials were first reported by Pelrine et al. \cite{pelrine1998electrostriction} and have since garnered interest as electromechanical transducers for a wide variety of applications, such as robotics, biomedical engineering, and energy harvesting (check, for example, \cite{jordi2010fish,anderson2010thin,shian2015dielectric,bar2001electroactive, brochu2010advances, kornbluh2012dielectric,bauer201425th,pei2016dielectric,wu2016energy, lee2022wearable}). To harness their full potential in engineering applications, robust constitutive models and computational schemes that can predict their complex electromechanical behavior under real-life loadings and stimuli are needed. Thus, work on electromechanical constitutive modeling has motivated research for decades \cite{toupin1956elastic,maugin1980method,maugin1992nonlinear,dorfmann2005nonlinear,bustamante2009nonlinear}. However, most of these models are limited to elastic dielectrics, which are dielectric elastomers that deform and polarize without dissipating energy. This idealization is not true in general for real-life loadings and applications, as dielectric elastomers are inherently dissipative solids that exhibit energy loss through viscous deformation and friction in their electric polarization process. Recognizing the dissipative nature of dielectric elastomers, many models focusing on the mechanical dissipation have also been proposed, but the majority of such models account only for mechanical dissipation and save for a few works most assume ideal dielectric behavior \cite{hong2011modeling,wang2016modeling,zhao2011nonequilibrium}.
This is because the electric relaxation time is much smaller than the mechanical relaxation time for most dielectric elastomers. However, accounting for electric dissipation in addition to mechanical dissipation becomes critical for several loading conditions such as in the presence of alternating electric potentials (check, for example, dielectric spectroscopy experiments on prestretched VHB 4910 specimen in \cite{qiang2012experimental}). In fact, the measured permittivity in such experiments becomes stretch dependent and was coined as \textit{apparent} permittivity in \cite{ghosh2021two}. Furthermore, an important use of dielectric elastomers is in composite material discovery in which dielectric elastomers, when filled with solid or liquid inclusions, lead to remarkable macroscopic or effective material properties \cite{yun2019liquid, ghosh2022elastomers}. The overall dissipative nature of these composites could be far more complex or pronounced than the electric and mechanical dissipation of the constituent elastomer \cite{ghosh2019homogenization,ghosh2021nonlinear, shrimali2023nonlinear}, thus highlighting the need for a constitutive model for the coupled electric and mechanical dissipative behavior of dielectric elastomers. A first attempt to propose such a comprehensive constitutive model was made in \cite{ghosh2021two}. The proposed model works for a prominent class of isotropic and incompressible dielectric elastomers that exhibit : a) non-Gaussian elasticity, b) deformation enhanced shear thinning viscosity, c) electrostriction and d) time- and deformation-dependent polarization. 

At this point, some work that has gone into addressing some of the challenges in FE implementation of constitutive models of dielectric elastomers needs to be acknowledged. These implementations have addressed challenges such as large deformation, electromechanical coupling, and the incorporation of viscoelastic effects (\cite{qu2012finite}, \cite{henann2013modeling}, see also \cite{stewart2024large} for an open-source implementation in \texttt{FEniCSx}) as well as efficient implementations of coupled-physics such as thermo-electro-viscoelasticity (\cite{mehnert2018numerical}), magnetorheological elastomers (\cite{dadgar2022large,rambausek2022computational}) and notably an efficient FE framework for coupled electromechanics \cite{kadapa2020robust}. To this end, the purpose of this paper is five-fold. Firstly, the ease of a robust FE implementation of the constitutive model proposed in \cite{ghosh2021two} is demonstrated. Second, various time-integration schemes are examined along with a re-examination of the time-integration scheme proposed in \cite{ghosh2021two}. Third, the FE framework is employed to describe the electromechanical behavior of the acrylate elastomer VHB 4910 and the results are compared with the experiments of Hossain et al. \cite{hossain2015comprehensive}. Fourth, it is demonstrated that such a FE framework is needed for finding material properties of the proposed model as boundary effects become important in such material characterization experiments invalidating the often made assumptions of biaxility/uniaxiality. This is because the sample geometry and the boundary conditions of the electro-mechanical experiments is such that the resulting electric and deformation fields are no longer homogeneous. This is also acknowledged in \cite{mehnert2021complete}. Finally, in a future work the FE solver will be used as a plug-in tool, within a larger framework that could involve using Machine Learning (ML), for data generation for automated constitutive model identification such as in \cite{martonova2024automated} or coupling with Deep Learning tools such as Neural ODEs~\cite{tacc2023data} for data-driven constitutive modeling modeling. 

The organization of the paper is as follows. In Section \ref{Sec:Problem}, the governing equations that take the form of an initial boundary value problem (IBVP) that explain the time-dependent dielectric response of a dielectric elastomer are outlined. The section begins with discussions on kinematics and the balance equations. Following this, the specific model for the prominent class of isotropic and incompressible dielectric elastomers as proposed in \cite{ghosh2021two} is explained. In order to deal with nearly or fully incompressible elastomers, it is more convenient to reformulate the governing equations into a hybrid set of governing equations in which a pressure field (in addition to the deformation field and the electric potential) is an additional unknown field. This hybrid form of the governing equations are described next, followed by the weak form of the equations. Finally, Section \ref{Sec:Problem} concludes with the time- and space- discretized forms of the IBVP that is to be implemented in a FE framework.

Section~\ref{Sec:NumericalScheme} is devoted to the study of explicit and implicit time integration schemes for the solution of the IBVP in the 1D setting. Specifically, the overall accuracy, stable-time-increment, and total-time-to-solution (TTS) for each of the schemes are studied. We note that implicit schemes are much better at handling the disparate time-scales exhibited by the electric and mechanical dissipation processes in elastomers. Hence these are better suited for solving the time-dependent response of dielectric elastomers exhibiting both mechanical and electric dissipation. Section~\ref{Sec:FEniCS-X} is devoted to the hybrid FE element formulation developed in Section~\ref{Sec:Problem} along with an implicit time-stepping algorithm based on Backward Euler discretization of time to solve an initial boundary value problem that mimics the corresponding experiments in \cite{hossain2015comprehensive}. Finally, Section~\ref{sec:Final_Comments} summarizes the findings, and presents concluding remarks along with possible extensions of the work in the future.

% \bsdelete{numerical schemes are attempted to solve the governing equations and the hybrid form of the governing equations via different numerical schemes.} The numerical scheme is a generalization of the numerical implementation of a non-linear viscoelastic formulation of elastomers presented in \cite{ghosh2021nonlinear,shrimali2023nonlinear}.

\section{The Problem}\label{Sec:Problem}
\subsection{Kinematics}
Consider a deformable and polarizable homogeneous solid that in its undeformed configuration (at time $t = 0$) is stress-free, unpolarized and occupies a bounded domain $\Omega_0\subset \mathbb{R}^3$, with boundary $\partial \Omega_0$ and unit outward normal $\bfN$. Material points are denoted by their initial position vector $\bfX\in \mathrm{\Omega}_0$. Due to applied stimuli, the solid deforms to its current configuration $\mathrm{\Omega}(t) \subset \mathbb{R}^3$ at a later time $t\in (0, T]$, such that the position vector $\bfX$ of a material point moves to a new position specified by
\begin{equation*}
\bfx=\bfy(\bfX,t),
\end{equation*}
where $\bfy$ is an invertible mapping from $\mathrm{\Omega}_0$ to $\mathrm{\Omega}(t)$. The deformed line elements $\text{d}\bfx \in \Omega(t)$ can be mapped to the undeformed material line elements $\text{d}\bfX \in \Omega_0$ through the deformation gradient and its determinant at $\bfX$ and $t$ as
\begin{equation*}
\bfF=\nabla\bfy(\bfX,t)=\dfrac{\partial \bfy}{\partial\bfX}(\bfX,t)\quad \text{such that}\quad J=\det\bfF>0.
\end{equation*}
\begin{figure}[h!]
    \centering
    \includegraphics[width=0.5\linewidth]{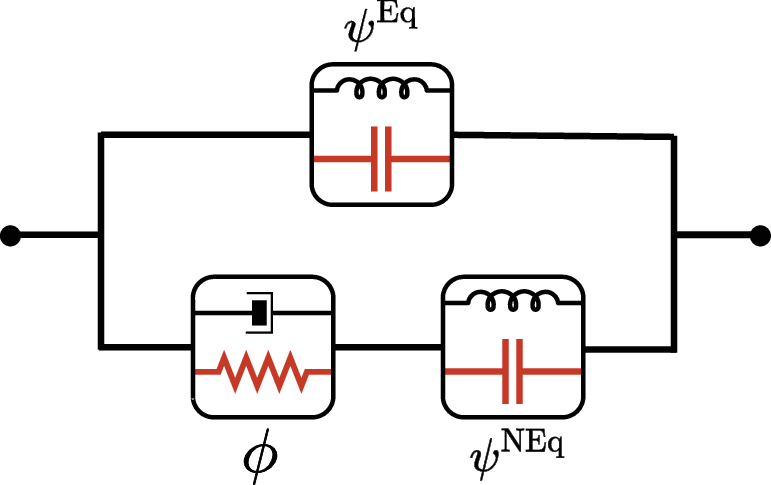}
    \caption{Rheological model of a dielectric elastomer. The elastic springs and the capacitor signify the elastomer's ability to store energy via mechanical deformation and via electric potential respectively. The dashpot and resistor on the other hand represent the elastomer's ability to dissipate energy via viscous dissipation and friction during the electric polarization process respectively. We refer the interested reader to ~\cite{ghosh2021two} for a detailed discussion of the model.}
    \label{fig:rheologicalModel}
\end{figure}
\subsection{Constitutive Equations}\label{Constitutive_Equations}
According to the two-potential framework discussed in \cite{ghosh2021two}, the constitutive behavior of such a deformable dielectric solid is characterized by two thermodynamic potentials -- a) the free-energy function: 
\begin{align}
\psi=\psi^{{\rm Eq}}(\bfF,\bfE)+\psi^{{\rm NEq}}(\bfF,\bfE,\bfF{\bfF^{v}}^{-1},\bfE-\bfE^v),\label{psi-gen}
\end{align}
and, b) the dissipation potential:  
\begin{align}
\phi=\phi(\bfF,\bfE,\bfF^v,\bfE^v,\dot{\bfF}^v,\dot{\bfE}^v,)\label{phi-gen}.
\end{align}

Here, $\psi$ denotes the total free-energy comprising of the energy storage in the dielectric elastomer at states of mechanical and electric equilibrium ($\psi^{{\rm Eq}}$) and the additional energy at non-equilibrium states ($\psi^{{\rm NEq}}$). Evidently, ($\psi^{{\rm Eq}}$) is a function of the deformation gradient $\bfF$ and the Lagrangian electric field $\bfE$ while, ($\psi^{{\rm NEq}}$) depends on $\bfF$, $\bfE$ and internal variables $\bfF^v$ and $\bfE^v$. These internal variables are measures of the dissipative parts of the deformation gradient and Lagrangian electric field respectively. The dissipation potential $\phi$ depends on $\bfF$, $\bfE$, $\bfF^v$, $\bfE^v$ and the rates of change of the interval variables denoted by $\dot{\bfF}^v=\partial\bfF^v(\bfX,t)/\partial t$ and $\dot{\bfE}^v=\partial\bfE^v(\bfX,t)/\partial t$\footnote{Here $\dot{()}$ denotes the material time derivative (i.e. derivative with respect to $t$ with $\bfX$ fixed)}.

Given these two potentials, the ``total''  first Piola-Kirchhoff stress tensor $\bfS$ (mechanical and Maxwell stress, \cite{dorfmann2005nonlinear}) and the Lagrangian electric displacement $\bfD$ at any material point $\bfX\in\mathrm{\Omega}_0$ and time $t\in[0,T]$ are given by the relations.
\begin{equation}
\bfS(\bfX,t)=\dfrac{\partial \psi^{{\rm Eq}}}{\partial\bfF}+\dfrac{\partial \psi^{{\rm NEq}}}{\partial\bfF}\label{S-gen}
\end{equation}
and
\begin{equation}
\bfD(\bfX,t)=-\dfrac{\partial \psi^{{\rm Eq}}}{\partial\bfE}-\dfrac{\partial \psi^{{\rm NEq}}}{\partial\bfE},\label{D-gen}
\end{equation}
where $\bfF^v(\bfX,t)$ and $\bfE^v(\bfX,t)$ are solutions of the evolution equations
\begin{equation}
    \begin{aligned}
        &\dfrac{\partial \psi^{{\rm NEq}}}{\partial\bfF^v}+\dfrac{\partial \phi}{\partial \dot\bfF^v}=0\\[10pt]
        &\dfrac{\partial \psi^{{\rm NEq}}}{\partial\bfE^v}+\dfrac{\partial \phi}{\partial \dot\bfE^v}=0 
    \end{aligned}
    ,\label{Evol-Eq-gen}
\end{equation}
with initial conditions $\bfF^v(\bfX,0)=\bfI$ and $\bfE^v(\bfX,0)=\bf0$.

This constitutive relationship is very general in its framework which allows specialized assumptions of $\psi$ and $\phi$ for a given dielectric elastomer to accurately estimate and predict its material behavior. For the remainder of the paper and as discussed in \cite{ghosh2021two}, the focus is on a prominent class of isotropic and incompressible dielectric elastomers which exhibit a) non-Gaussian elasticity, b) deformation enhanced shear thinning viscosity, and c) time- and deformation-dependent polarization and electrostriction. We choose the specific form of the functions $\psi^{{\rm Eq}}$, $\psi^{{\rm NEq}}$ and $\phi$ in this work to be the same as discussed in \cite{ghosh2021two}. To this end, the equilibrium free-energy $\psi^{{\rm Eq}}$ is given by 
\begin{equation}
    \psi^{{\rm Eq}}
    =
    \begin{cases}
    \sum\limits_{r=1}^2\,\dfrac{3^{1-\alpha_{r}}}{2\alpha_{r}}\mu_{r} \left[I_{1}^{\alpha_{r}} -3^{\alpha_{r}}\right]+\dfrac{m_{\texttt{K}}-\varepsilon}{2}I_4-\dfrac{m_{\texttt{K}}}{2}I_5\quad & {\rm if} \quad J= 1 \\[10pt]
    +\infty                                        
    & {\rm otherwise} ,
    \end{cases}
    \label{Proposed model-Eq}
\end{equation}
where $I_1$, $I_4$, $I_5$ are the invariants
\begin{equation*}
I_1={\rm tr}\,\bfC,\quad I_4=\bfE\cdot\bfE,\quad I_5=\bfE\cdot\bfC^{-1}\bfE,
\end{equation*}
$\bfC=\bfF^T\bfF$ is the right Cauchy-Green deformation tensor, and $\mu_r$, $\alpha_r$ ($r=1,2$) are the material parameters describing the elasticity of the dielectric elastomer, while the material parameters $\varepsilon$ and $m_{\texttt{K}}$ denote, respectively, its initial permittivity and electrostriction coefficient at states of mechanical and electric equilibrium.

Analogous to its equilibrium counterpart the non-equilibrium free-energy function $\psi^{{\rm NEq}}$ is given by 
\begin{equation}
    \psi^{{\rm NEq}}=
    \begin{cases}
        \sum\limits_{r=1}^2\dfrac{3^{1-\beta_r}}{2\beta_r}\nu_{r} \left[{I^{e}_{1}}^{\beta_r} -3^{\beta_r}\right]+\dfrac{n_{\texttt{K}}-\epsilon}{2}I^e_4-\dfrac{n_{\texttt{K}}}{2}\mathcal{I}^e_5 & \text{if } J^e= 1 \\[10pt]
        +\infty & \text{otherwise}
    \end{cases} 
    \label{Proposed model-NEq}
\end{equation}
where,
\begin{align}
\begin{aligned}
    I^e_1&={\rm tr}\,\bfC^e=\bfC\cdot{\bfC^v}^{-1} \\[10pt]
    J^e&=\det\bfF^e= \dfrac{J}{J^v}\\[10pt]
    I^e_4&=\bfE^e\cdot\bfE^e=(\bfE-\bfE^v)\cdot(\bfE-\bfE^v)\\[10pt]
    \mathcal{I}^e_5&=\bfE^e\cdot\bfC^{-1}\bfE^e=(\bfE-\bfE^v)\cdot\bfC^{-1}
    (\bfE-\bfE^v) \\[10pt]
    \bfC^e&={\bfF^e}^T\bfF^e, \bfC^v={\bfF^{v}}^{T}\bfF^{v}, J^v=\det\bfF^v
\end{aligned}
\label{Ies}
\end{align}
and where  $\nu_1$, $\beta_1$, $\nu_2$, $\beta_2$, $\epsilon$, $n_{\texttt{K}}$ are the six material parameters analogous to $\mu_1$, $\alpha_1$, $\mu_2$, $\alpha_2$, $\varepsilon$, $m_{\texttt{K}}$ in the equilibrium branch (\ref{Proposed model-Eq}). 

Finally, the dissipation potential $\phi$ is prescribed as
\begin{align}
\phi=&\dfrac{1}{2}\dot{\bfF}^v{\bfF^v}^{-1}\cdot\left[\Atan\left(\bfF,\bfF^e\right)\dot{\bfF}^v{\bfF^v}^{-1}\right]+\dfrac{1}{2}\bfF^{-T}\dot{\bfE}^v\cdot\left[\Btan\left(\bfF,\bfE,\bfE^e\right)\bfF^{-T}\dot{\bfE}^v\right]\label{Proposed model-phi}
\end{align}
with
\begin{align}
\Atan(\bfF, \bfF^{e})=&  2\eta_{\texttt{K}}(I_1^e,I_2^e,I_1^v)\mathcal\Ktan+3\eta_{\texttt{J}}\Jtan
\label{Proposed model-A}
\end{align}
and
\begin{align}
\Btan\left(\bfF,\bfE,\bfE^e\right)= \zeta\bfI,\label{Proposed model-B}
\end{align}
where $\bfI$ is the identity tensor, and $\Ktan$ and $\Jtan$ stand for the standard shear and hydrostatic orthogonal projection tensors
\begin{align}
\mathcal{K}_{ijkl}=\frac{1}{2}\left[\delta_{ik}\delta_{jl}+\delta_{il}\delta_{jk}-\frac{2}{3}\delta_{ij}\delta_{kl}\right],
\;\mathcal{J}_{ijkl}=\frac{1}{3}\delta_{ij}\delta_{kl}.\label{K-J}
\end{align}
In (\ref{Proposed model-A})--(\ref{Proposed model-B}), $\eta_{\texttt{K}}$, $\eta_{\texttt{J}}$, and $\zeta$ are material functions/parameters that describe the non-linear shear-thinning type viscosity and the polarization friction of the dielectric elastomer. For the dielectric elastomer of interest, the two viscosity functions $\eta_{\texttt{K}}$ and $\eta_{\texttt{J}}$ are given by
\begin{equation*}
\eta_{\texttt{K}}(I_1^e,I_2^e,I_1^v)=\eta_{\infty}+\frac{\eta_{0}-\eta_{\infty}+K_{1}\left[{I_{1}^{v}}^{\gamma_{1}}-3^{\gamma_{1}}\right]}{1+\left(K_{2}\, J_2^{{\rm NEq}}\right)^{\gamma_{2}}} %\label{Proposed model-etaK}
\end{equation*}
with
\begin{equation*}
J_2^{{\rm NEq}}=\left(\dfrac{I_1^{e\,2}}{3}-I_2^{e}\right)\left(\sum_{r=1}^2 3^{1-\beta_{r}}\nu_{r} I_{1}^{e\,\beta_{r}-1} \right)^{2} %\label{Proposed model-etaK}
\end{equation*}
and
\begin{equation}
\eta_{\texttt{J}}=+\infty,
\label{Proposed model-etaJ}
\end{equation}
where $I_1^e$ is given by (\ref{Ies})$_1$, and
\begin{align*}
% \left{ \begin{array}{ll}
% I^e_2&=\dfrac{1}{2}\left[({\rm tr}\,\bfC^e)^2-{\rm tr}\,{\bfC^e}^2\right]\\[10pt]
% &=\dfrac{1}{2}\left[(\bfC\cdot{\bfC^v}^{-1})^2-{\bfC^v}^{-1}\bfC\cdot\bfC{\bfC^v}^{-1}\right]\\[10pt]
% I_1^v&={\rm tr}\,\bfC^v \end{array}
%             \right. ,%\label{Ievs}
    \begin{aligned}
        I^e_2&=\dfrac{1}{2}\left[({\rm tr}\,\bfC^e)^2-{\rm tr}\,{\bfC^e}^2\right] \\[10pt]
        &=\dfrac{1}{2}\left[(\bfC\cdot{\bfC^v}^{-1})^2-{\bfC^v}^{-1}\bfC\cdot\bfC{\bfC^v}^{-1}\right]\\[10pt]
        I_1^v&={\rm tr}\,\bfC^v
    \end{aligned},
\end{align*}
where $\eta_0>\eta_\infty\geq 0$, $\gamma_1\geq 0$, $\gamma_2\geq 0$, $K_1\geq 0$, $K_2\geq 0$ are material constants. 

With these constitutive prescriptions, first Piola-Kirchhoff stress tensor $\bfS$ and the Lagrangian electric displacement $\bfD$ at any material point $\bfX\in\mathrm{\Omega}_0$ and time $t\in[0,T]$ can be calculated using (\ref{S-gen})-(\ref{D-gen}), namely
\begin{align}
\bfS =& \left[\sum\limits_{r=1}^2 3^{1-\alpha_{r}}\mu_{r} I_{1}^{\alpha_{r}-1} \right] \bfF -p\, \bfF^{-T}+\left[\sum\limits_{r=1}^2 3^{1-\beta_{r}} \nu_{r} (\bfC \cdot {\bfC ^{v}}^{-1} )^{\beta_{r}-1} \right] \bfF {\bfC ^{v}}^{-1}+\nonumber\\
& m_{\texttt{K}}\bfF^{-T}\bfE\otimes \bfF^{-1}\bfF^{-T}\bfE +n_{\texttt{K}}\bfF^{-T}(\bfE-\bfE^v)\otimes \bfF^{-1}\bfF^{-T}(\bfE-\bfE^v)
\label{Proposed CR- P-K stress}
\end{align}
and
\begin{align}
\bfD=&\left(\varepsilon-m_{\texttt{K}}\right)\bfE+m_{\texttt{K}}\bfF^{-1}\bfF^{-T}\bfE+\left(\epsilon-n_{\texttt{K}}\right)(\bfE-\bfE^v)+n_{\texttt{K}}\bfF^{-1}\bfF^{-T}(\bfE-\bfE^v),\label{Proposed D}
\end{align}
where $p$ is the hydrostatic pressure associated with the incompressibility constraint $J=1$, and where $\bfC ^{v}$ and $\bfE^v$ are implicitly defined by the evolution equations
\begin{align}
    \begin{aligned}
        &\dot{\bfC} ^{v}=\frac{\sum\limits_{r=1}^2 3^{1-\beta_{r}} \nu_{r} (\bfC \cdot {\bfC ^{v}}^{-1} )^{\beta_{r}-1}}{\eta_{\texttt{K}}(I_1^e,I_2^e,I_1^v)} \left[\bfC - \frac{1}{3} (\bfC \cdot {\bfC ^{v}}^{-1}) \bfC^v\right] \\[10pt]
        &\bfC^v(\bfX,0)=\bfI
    \end{aligned},
\label{Evol-Eq-Cv}
\end{align}
and
\begin{align}
    \begin{aligned}
        &\dot{\bfE} ^{v}
        =-\left(\dfrac{n_{\texttt{K}}}{\zeta}\bfI+\dfrac{\epsilon-n_{\texttt{K}}}{\zeta} \bfC\right)(\bfE-\bfE^v) \\[10pt]
        &\bfE^v(\bfX,0)={\bf0}
    \end{aligned}.
    \label{Evol-Eq-Ev}
\end{align}

\subsection{Balance Equations}

In the absence of inertia and in the context of electro-quasi-statics, the equations governing the balance of linear and angular momenta and electrostatics in the undeformed configuration are given by
\begin{equation}\label{BM}
    {\rm Div}\,\bfS+\mathbf{f}=\textbf{0}\quad {\rm and }\quad \bfS\bfF^T=\bfF\bfS^T,\quad (\bfX,t)\in\mathrm{\Omega}_0\times [0,T]
\end{equation}
and 
\begin{equation}\label{Maxwell}
    {\rm Div}\,\bfD=Q\quad {\rm and }\quad {\rm Curl}\,\bfE=\textbf{0},\quad (\bfX,t)\in \mathbb{R}^3\times [0,T]
\end{equation}
respectively. Here, $\mathbf{f}$ are the body forces per unit volume and $Q$ are the space charges per unit volume. 

While, the balance of angular momentum (\ref{BM})$_2$ is satisfied by the material frame indifference of the constitutive prescriptions \eqref{Proposed model-Eq}, \eqref{Proposed model-NEq} and \eqref{Proposed model-phi}, Faraday's law (\ref{Maxwell})$_2$ can directly satisfied by introducing a scalar potential $\varphi$ such that $\bfE=-\nabla\varphi(\bfX,t)$. Furthermore, with the introduction of the scalar $\varphi(\bfX,t)$, (\ref{Maxwell})$_2$ can now be solved over the domain of the material $\mathrm{\Omega}_0$ instead of over all $\mathbb{R}^3$. Along with the constitutive model \eqref{Proposed CR- P-K stress}-\eqref{Evol-Eq-Ev}, the governing equations for the solid then reduce to the following system of coupled initial boundary-value problems (IBVPs):
%
% \begin{widetext}
%
\begin{align}
\hspace*{-2ex}&\left\{
    \begin{aligned}
        {\rm Div}\left[\bfS\left(\nabla\bfy,-\nabla\varphi,\bfC^v,\bfE^v\right)\right]+\mathbf{f}(\bfX,t)&={\bf0}, \quad (\bfX,t)\in \mathrm{\Omega}_0\times[0,T]\\[10pt]
        \det\nabla\bfy(\bfX,t)&>0,\quad (\bfX,t)\in\mathrm{\Omega}_0\times[0,T]\\[10pt]
        \bfy(\bfX,t)&=\overline{\bfy}(\bfX,t),\quad (\bfX,t)\in\partial\mathrm{\Omega}_0^{\mathcal{D}}\times[0,T]\\[10pt]
        \left[\bfS\left(\nabla\bfy,-\nabla\varphi,\bfC^v,\bfE^v\right)-\bfS_M\right]\bfN&=\overline{\textbf{t}}(\bfX,t),\; (\bfX,t)\in\partial \mathrm{\Omega}^{\mathcal{N}}_0\times[0,T]
    \end{aligned}
    \right. \label{BVP-F}\\[10pt]
    &{\rm and} \nonumber \\[10pt]
    &\left\{
        \begin{aligned}
        {\rm Div}\left[\bfD\left(\nabla\bfy,-\nabla\varphi,\bfC^v,\bfE^v\right)\right]&=-Q(\bfX,t), \quad (\bfX,t)\in \mathrm{\Omega}_0\times[0,T]\\[10pt]
        \varphi(\bfX,t)&=\overline{\Phi}(\bfX,t),\quad (\bfX,t)\in\partial\mathrm{\Omega}_0^{\mathcal{D}_{\phi}}\times[0,T]\\[10pt]
        \left[
        \bfD\left(\nabla\bfy , -\nabla\varphi, \bfC^v, \bfE^v \right)\right]\cdot\bfN&=\sigma(\bfX,t)-\overline{\bfD}(\bfX,t)\cdot\bfN, \;(\bfX,t)\in\partial \mathrm{\Omega}^{\mathcal{N}_\phi}_0\times[0,T]
        \end{aligned}
    \right.\label{BVP-E}
% &\left{\begin{array}{ll}
%  {\rm Div}\left[\bfS\left(\nabla\bfy,-\nabla\varphi,\bfC^v,\bfE^v\right)\right]+\mathbf{f}(\bfX,t)={\bf0}, \quad (\bfX,t)\in \mathrm{\Omega}_0\times[0,T]\\[10pt]
% \det\nabla\bfy(\bfX,t)>0,\quad (\bfX,t)\in\mathrm{\Omega}_0\times[0,T]\\[10pt]
% \bfy(\bfX,t)=\overline{\bfy}(\bfX,t),\quad (\bfX,t)\in\partial  \mathrm{\Omega}_0^{\mathcal{D}}\times[0,T]\\[10pt]
%  \left[\bfS\left(\nabla\bfy,-\nabla\varphi,\bfC^v,\bfE^v\right)-\bfS_M\right]\bfN=\overline{\textbf{t}}(\bfX,t),\; (\bfX,t)\in\partial \mathrm{\Omega}_0^{\mathcal{N}}\times[0,T]\end{array}\right.  \label{BVP-F}\\[10pt]
% &{\rm and} \nonumber \\[10pt]
% &\left{\begin{array}{ll}
%      {\rm Div}\left[\bfD\left(\nabla\bfy,-\nabla\varphi,\bfC^v,\bfE^v\right)\right]=-Q(\bfX,t), \quad (\bfX,t)\in \mathrm{\Omega}_0\times[0,T]\\[10pt]
%      \varphi(\bfX,t)=\overline{\Phi}(\bfX,t),\quad (\bfX,t)\in\partial  \mathrm{\Omega}_0^{\mathcal{D}_{\phi}}\times[0,T]\\[10pt]
%      \left[
%         \bfD\left(\nabla\bfy , -\nabla\varphi, \bfC^v, \bfE^v \right)\right]\cdot\bfN=\sigma(\bfX,t)-\overline{\bfD}(\bfX,t)\cdot\bfN, \;(\bfX,t)\in\partial \mathrm{\Omega}^{\mathcal{N}_\phi}_0\times[0,T]
%  \end{array}\right.\label{BVP-E}
\end{align}
%
% \end{widetext}
%
together with the coupled system of evolution equations 
% \eqref{Evol-Eq-Cv}-\eqref{Evol-Eq-Ev}:
%
\begin{align}
% \left{ \begin{array}{ll} \dot{\bfC} ^{v}=\dfrac{\sum\limits_{r=1}^2 3^{1-\beta_{r}} \nu_{r} (\bfC \cdot {\bfC ^{v}}^{-1} )^{\beta_{r}-1}}{\eta_{\texttt{K}}(I_1^e,I_2^e,I_1^v)} \left[\nabla\bfy^T \nabla\bfy - \frac{1}{3} (\nabla\bfy^T \nabla\bfy \cdot {\bfC ^{v}}^{-1}) \bfC^v\right]  = \mathbf{G}(\nabla\bfy^T \nabla\bfy,\bfC ^{v})\\[10pt]
% \bfC^v(\bfX,0)=\bfI
% \end{array}
%             \right.
% \label{Evol-Eq-Cv2}
    \begin{aligned}
    &\dot{\bfC}^{v}=\dfrac{\sum\limits_{r=1}^2 3^{1-\beta_{r}} \nu_{r} (\bfC \cdot {\bfC ^{v}}^{-1} )^{\beta_{r}-1}}{\eta_{\texttt{K}}(I_1^e,I_2^e,I_1^v)} \left[\nabla\bfy^T \nabla\bfy - \frac{1}{3} (\nabla\bfy^T \nabla\bfy \cdot {\bfC ^{v}}^{-1}) \bfC^v\right] 
    = \mathbf{G}(\nabla\bfy^T \nabla\bfy,\bfC ^{v})\\[10pt]
    &\bfC^v(\bfX,0)=\bfI
    \end{aligned}\label{Evol-Eq-Cv2}
\end{align}
and
\begin{align}
    \begin{aligned}
        &\dot{\bfE}^{v}=-\left(\dfrac{n_{\texttt{K}}}{\zeta}\bfI+\dfrac{\epsilon-n_{\texttt{K}}}{\zeta} \nabla\bfy^T \nabla\bfy\right)(-\nabla \varphi-\bfE^v) = \mathbf{H}(\nabla\bfy^T \nabla\bfy, \nabla \varphi,\bf E^v) \\[10pt]
        &\bfE^v(\bfX,0)={\bf0}
    \end{aligned}.\label{Evol-Eq-Ev2}
\end{align}
In (\ref{BVP-F}), $\overline{\textbf{y}}(\bfX,t)$ are the Dirichlet boundary conditions for the displacement field $\bfy(\bfX,t)$ and $\overline{\textbf{t}}(\bfX,t)$ are the surface tractions per unit {undeformed} area which is is equal to the jump of tractions due to the presence of Maxwell stress $\bfS_{M}$ outside of the solid\footnote{In the case when the solid is surrounded by air, $\bfS_{M}=\obfF^{-T}\overline{\bfE}\otimes\overline{\bfD}-\dfrac{\overline{J}\varepsilon_0}{2}\left(\bfF^{-T}\overline{\bfE}\cdot\obfF^{-T}\overline{\bfE}\right)\obfF^{-T}$.}. In (\ref{BVP-E}), $\overline{\Phi}$ is the Dirichlet boundary condition on $\varphi(\bfX,t)$ and $\sigma(\bfX,t)$ are the surface charges per unit {undeformed} area satisfying the jump of Dielectric displacement fields\footnote{If the solid is surrounded by air, $\overline{\bfD}=\varepsilon_0 \overline{J}\,\obfF^{-1}\obfF^{-T}\overline{\bfE}$.}. 

\subsection{Hybrid Formulation}

As stated in Section \ref{Constitutive_Equations}, the focus of this work is on isotropic incompressible dielectric elastomers which is also evident from the constitutive model in \eqref{Proposed model-Eq}, \eqref{Proposed model-NEq} and \eqref{Proposed model-etaJ}. To deal with incompressible or {nearly}-incompressible elastomers, equations \eqref{BVP-F}-\eqref{Evol-Eq-Ev2} cannot be used directly. Instead a reformulated equivalent set of equations needs to be derived, wherein in addition to the unknown displacements and electric-potentials, a pressure field needs to be solved as well. This hinges on introduction of an appropriate Legendre transform as is done in the simpler settings of elasticity \cite{chi2015polygonal}, elastic dielectric elastomers \cite{lefevre2017nonlinear}, nonlinear viscoelasticity \cite{ghosh2021nonlinear,shrimali2023nonlinear} and elasticity with interfacial mechanics \cite{ghosh2022elastomers}.

% Invoking the compressible versions of the consitutive model functions (\ref{Proposed model-Eq}) - (\ref{Proposed model-NEq}) given by
% %
% ,
% %
% now consider the function
{To this end, consider a \textit{compressible} dielectric elastomer} characterized by the total stored-energy function
\begin{equation}
    W(I_1,I_4,I_5,J,I^e_1,I^e_4,\mathcal{I}^e_5) = \psi^{{\rm Eq}}(I_1,I_4,I_5) + \psi^{{\rm NEq}}(I^e_1,I^e_4,\mathcal{I}^e_5) + \sum_{r=1}^2 \mu_{r} \ln (J)+\frac{\kappa}{2}(J-1)^2
\end{equation}
along with its partial Legendre transform
\begin{align}
    \psi^{\star}(I_1,I_4,I_5,I^e_1,I^e_4,\mathcal{I}^e_5,p) 
    &= \max_{J}\left\{
    p(J-1)-W(I_1,I_4,I_5,J,I^e_1,I^e_4,\mathcal{I}^e_5)
    \right\} \nonumber\\
    & = p(J^{\star}-1)-W(I_1,I_4,I_5,J^{\star},I^e_1,I^e_4,\mathcal{I}^e_5),\label{Legendre-W-Dual}
\end{align}
with $J^{\star} = \dfrac{(1+p/\kappa) + \sqrt{(1+p/\kappa)^2 + 4(\rm \sum_{r=1}^2 \mu_{r}/\kappa)}}{2}$. Furthermore, since $W$ is convex in $J$, it follows that
\begin{align}
    \hspace*{-2ex}\psi(I_1,I_4,I_5,J,I^e_1,I^e_4,\mathcal{I}^e_5) 
    &= {\left( {\psi^{\star}} \right)}^{\star}(I_1,I_4,I_5,J,I^e_1,I^e_4,\mathcal{I}^e_5,p) \nonumber\\
    & = \max_{p}\left\{p(J-1)-\psi^{\star}(I_1,I_4,I_5,J^{\star},I^e_1,I^e_4,\mathcal{I}^e_5)\right\} \nonumber\\
    & = \max_{p}\left\{p(J-J^{\star}) + \psi^{{\rm Eq}}(I_1,I_4,I_5) + \psi^{{\rm NEq}}(I^e_1,I^e_4,\mathcal{I}^e_5) 
    + \sum_{r=1}^2 \mu_{r} \ln (J^{\star})+\frac{\kappa}{2}(J^{\star}-1)^2\right\},\label{S_gen_hybrid}
\end{align}
which in turn allows the first Piola-Kirchhoff stress tensor (\ref{S-gen}) to be re-written as 
\begin{align}
    % \left{ \begin{array}{ll} \bfS^{\rm{hybrid}}(\bfX,t)=\dfrac{\partial \psi}{\partial\bfF} &= \dfrac{\partial \psi^{{\rm Eq}}}{\partial\bfF}+\dfrac{\partial \psi^{{\rm NEq}}}{\partial\bfF} + pJ \bfF^{-T}\\[10pt]
    % \rm{with} \\ [10pt]
    % \dfrac{\partial \psi}{\partial p}=0 \Rightarrow J=1+\dfrac{\partial \psi^{\star}}{\partial p}.
    % \end{array}
    %             \right.
    \begin{aligned}
        \bfS^{\rm{hybrid}}(\bfX,t) &= \dfrac{\partial \psi}{\partial\bfF} \\
        &= \dfrac{\partial \psi^{{\rm Eq}}}{\partial\bfF}+\dfrac{\partial \psi^{{\rm NEq}}}{\partial\bfF} + pJ \bfF^{-T} 
    \end{aligned}
    \label{S_gen_hybrid_2}
\end{align}
with
\begin{equation}
    \dfrac{\partial \psi}{\partial p} = 0 \Rightarrow J=1+\dfrac{\partial \psi^{\star}}{\partial p}.\label{eq:J_dW_star_dp}
\end{equation}
Combining \eqref{S_gen_hybrid_2} and \eqref{eq:J_dW_star_dp}, the hybrid form of the governing equations \eqref{BVP-F}-\eqref{Evol-Eq-Ev2} can be re-written as
%
% \begin{widetext}
%
\begin{align}
\hspace*{-2ex}&\left\{\begin{array}{ll}
 {\rm Div}\left[\bfS\left(\nabla\bfy ,-\nabla\varphi,\bfC^v,\bfE^v\right) + p J \nabla \bfy^{-T}\right] + \mathbf{f}(\bfX,t)={\bf 0}, \quad (\bfX , t)\in \mathrm{\Omega}_0\times[0,T]\\[10pt]
\det\nabla\bfy(\bfX,t) =  \dfrac{1 + p/\kappa+\sqrt{4\rm \sum_{r=1}^2\mu_r/\kappa + (1+p/\kappa)^2}}{2},\quad (\bfX, t)\in\mathrm{\Omega}_0\times[0,T]\\[10pt]
\bfy(\bfX, t)=\overline{\bfy}(\bfX, t),\quad (\bfX, t)\in\partial  \mathrm{\Omega}_0^{\mathcal{D}}\times[0,T]\\[10pt]
 \left[\bfS \left( \nabla\bfy, -\nabla\varphi,\bfC^v,\bfE^v\right)+ p J \bfF^{-T} -\bfS_M \right]\bfN
 =
 \overline{\mathbf{t}}(\bfX, t),\; (\bfX, t)\in\partial \mathrm{\Omega}_0^{\mathcal{N}}\times[0,T]\end{array}\right.\label{BVP-F-hybrid}\\[10pt]
&{\rm and} \nonumber \\[10pt]
&\left\{\begin{array}{ll}
 {\rm Div}\left[\bfD\left(\nabla\bfy, -\nabla\varphi, \bfC^v,\bfE^v\right)\right]=-Q(\bfX,t), \quad (\bfX, t)\in \mathrm{\Omega}_0\times[0,T]\\[10pt]
 \varphi(\bfX,t)=\overline{\Phi}(\bfX,t),\quad (\bfX,t)\in\partial  \mathrm{\Omega}_0^{\mathcal{D}_{\phi}}\times[0, T]\\[10pt]
 \left[\bfD\left(\nabla\bfy,-\nabla\varphi,\bfC^v,\bfE^v\right)\right]\cdot\bfN=\sigma(\bfX,t)-\overline{\bfD}(\bfX,t)\cdot\bfN, \;(\bfX, t)\in\partial \mathrm{\Omega}^{{\mathcal{N}_\phi}}_0\times[0,T]\end{array}\right.  \label{BVP-E-hybrid}
\end{align}
%
% \end{widetext}
%
together with the coupled system of evolution equations \eqref{Evol-Eq-Cv2}-\eqref{Evol-Eq-Ev2}, namely,
\begin{align}
\left. \begin{array}{ll} \dot{\bfC} ^{v}=\mathbf{G}(\nabla\bfy^T \nabla\bfy,\bfC ^{v}) \\[10pt]
\bfC^v(\bfX,0)=\bfI
\end{array}
            \right.
            \label{Evol-Eq-Cv3}
\end{align} 
and
\begin{align}
\left. \begin{array}{ll} \dot{\bfE} ^{v}=\mathbf{H}(\nabla\bfy^T \nabla\bfy, \nabla \varphi,\bf E^v) \\[10pt]
\bfE^v(\bfX,0)={\bf 0}
\end{array}
            \right. .\label{Evol-Eq-Ev3}
\end{align}
%
% \bsdelete{The weak form of these hybrid equations are solved via time and space discretizations as detailed in Section \ref{Sec:FEniCS-X}. {We now proceed to the solution of the weak form of these hybrid equations. Specifically we outline the corresponding spatial (FE) and temporal (FD) discretizations.} {active vs passive voice or a mix of both; discuss with KG}}

\subsection{Time and Space Discretized Weak Form} \label{Sec:Time_Space_Diretized_Weak_Form}
\subsubsection{Weak Form}

The weak form of the hybrid equations \eqref{BVP-F-hybrid}-\eqref{BVP-E-hybrid} amounts to finding $\bfy(\bfX,t)\in \mathcal{U}$, $\phi(\bfX,t)\in \mathcal{V}$ and $p(\bfX,t)\in\mathcal{P}$ such that 
\begin{align}
    \hspace*{-2ex}&\left.\begin{array}{ll}
     \displaystyle\int_{\Omega_0}\left[\bfS\left(\nabla\bfy,-\nabla\varphi,\bfC^v,\bfE^v\right) + pJ \nabla \bfy^{-T}\right] \cdot \nabla \mathbf{v} \ \mathrm{d} \bfX=\\[10pt]
     \qquad \qquad \displaystyle\int_{\Omega_0}\mathbf{f}(\bfX,t)\cdot \mathbf{v} \ \mathrm{d} \bfX+ \displaystyle\int_{\partial \Omega_0^{\mathcal{N}}}\overline{\textbf{t}}(\bfX,t)\cdot \mathbf{v} \ \mathrm{d} \bfX, \quad \forall \mathbf{v} \in \mathcal{Y}_0, \textrm{$t$} \in [0,T]\\[10pt]
    \displaystyle\int_{\Omega_0}\left(\det\nabla\bfy(\bfX,t) -\dfrac{1+p/\kappa+\sqrt{4\rm \sum_{r=1}^2\mu_r/\kappa + (1+\textrm{$p$}/\kappa)^2}}{2}\right)q \ \mathrm{d} \bfX = 0,\quad \forall \textrm{$q$}\in \mathcal{P}_0, \textrm{$t$}\in [0,T]\\[10pt]
     \displaystyle\int_{\Omega_0}{\rm Div}\left[\bfD\left(\nabla\bfy,-\nabla\varphi,\bfC^v,\bfE^v\right)\right]\cdot \phi \ \mathrm{d} \bfX = \\[10pt]
     \qquad \qquad \displaystyle\int_{\Omega_0}Q(\bfX,t)\phi \ \mathrm{d} \bfX + \displaystyle\int_{\partial \mathrm{\Omega}^{\mathcal{N}_\phi}_0}\left(\sigma(\bfX,t)-\overline{\bfD}(\bfX,t)\cdot\bfN\right) \ \mathrm{d} \bfX,\quad \forall \textrm{$\phi$}\in \mathcal{V}_0, \textrm{$t$}\in [0,T]\end{array}\right.  \label{BVP-F-hybrid-weak}
\end{align}
where, $\bfC^v(\bfX)$ and $\bfE^v(\bfX,t)$ are defined by \eqref{Evol-Eq-Cv3}-\eqref{Evol-Eq-Ev3}, and where $\mathcal{Y}$, $\mathcal{V}$ and $\mathcal{P}$ are sufficiently large sets of admissible deformations $\bfy(\bfX,t)$, electric potentials $\varphi(\bfX,t)$ and pressures $p(\bfX,t)$ respectively. Furthermore, $\mathcal{Y}_0$, $\mathcal{V}_0$ and $\mathcal{P}_0$ are sufficiently large sets of test functions $v$, $\phi$ and $q$, respectively (see, e.g. \cite{lefevre2017nonlinear,tian_2012} for a description of these sets).\footnote{
\begin{align*}
    \begin{cases}
       \mathcal{U} = \{
           \bfy:\bfy(\bfX,t) = \overline{\bfy}(\bfX,t),\ \bfX\in \partial \Omega^{\mathcal{D}}
       \} \quad \qquad 
       \mathcal{U}_0 = \{
           \bfy:\bfy(\bfX,t) = 0,\ \bfX\in \partial \Omega^{\mathcal{D}}
       \} \\
       \mathcal{V} = \{
           \varphi:\varphi(\bfX,t) = \overline{\Phi}(\bfX,t),\ \bfX\in \partial \Omega_0^{\mathcal{D}_{\phi}}
       \} \qquad \mathcal{V}_0 = \{
           \varphi:\varphi(\bfX,t) = 0,\ \bfX\in \partial \Omega_0^{\mathcal{D}_\phi}
       \}
    \end{cases}
\end{align*}
}

\subsubsection{Time Discretization}
Consider the division of time interval $[0,T]$ into discrete times $t_{n+1} \in \{0=t_0,t_1,t_2,...,t_M=T\}$. Using the notations: $\bfy_{n+1} = \bfy(\bfX,t_{n+1})$, $\nabla \bfy_{n+1} = \nabla \bfy(\bfX,t_{n+1})$, $\varphi_{n+1} = \varphi(\bfX,t_{n+1})$, $\nabla \varphi_{n+1} = \nabla \varphi(\bfX,t_{n+1})$, $\bfC^v_{n+1} = \bfC^v(\bfX,t_{n+1})$ and $\bfE^v_{n+1}= \bfE^v(\bfX,t_{n+1})$, and similarly for any time-dependent field, the weak form for any discrete time $t_{n+1}$ then becomes
\begin{align}
    \hspace*{-2ex}&\left\{\begin{array}{ll}
     \displaystyle\int_{\Omega_0}
     \left[\bfS_{n+1}
     \left(\nabla\bfy_{n+1},-\nabla\varphi_{n+1},\bfC^v_{n+1},\bfE^v_{n+1}\right) + p_{n+1}J_{n+1} \nabla \bfy_{n+1}^{-T}\right] \cdot \nabla \mathbf{v} \ \mathrm{d} \bfX=\\[10pt]
     \qquad \qquad \displaystyle\int_{\Omega_0}\mathbf{f}_{n+1}(\bfX,t)\cdot \mathbf{v} \ \mathrm{d} \bfX+ \displaystyle\int_{\partial \Omega_0^{\mathcal{N}}}\overline{\textbf{t}}_{n+1}(\bfX,t)\cdot \mathbf{v} \ \mathrm{d} \bfX, \quad \forall \mathbf{v} \in \mathcal{Y}_0\\[10pt]
    \displaystyle\int_{\Omega_0}\left(\det\nabla\bfy_{n+1}(\bfX,t) -\dfrac{1+p_{n+1}/\kappa+\sqrt{4\rm \sum_{r=1}^2\mu_r/\kappa + (1+p_{n+1}/\kappa)^2}}{2}\right)q \ \mathrm{d} \bfX = 0,\quad \forall \textrm{$q$}\in \mathcal{P}_0\\[10pt]
     \displaystyle\int_{\Omega_0}{\rm Div}\left[\bfD_{n+1}\left(\nabla\bfy_{n+1},-\nabla\varphi_{n+1},\bfC_{n+1}^v,\bfE_{n+1}^v\right)\right]\cdot \phi \ \mathrm{d} \bfX = \\[10pt]
     \qquad \qquad \displaystyle\int_{\Omega_0}Q_{n+1}(\bfX,t)\phi \ \mathrm{d} \bfX + \displaystyle\int_{\partial \mathrm{\Omega}^{\mathcal{N}_\phi}_0}\left(\sigma_{n+1}(\bfX,t)-\overline{\bfD}_{n+1}(\bfX,t)\cdot\bfN\right) \ \mathrm{d} \bfX,\quad \forall \textrm{$\phi$}\in \mathcal{V}_0,\end{array}\right.  \label{BVP-F-hybrid-weak_disc}
\end{align}
and the time-discretized evolution equations:
\begin{align}
    \dot{\bfC}_{n+1} ^{v}=\mathbf{G}(\nabla\bfy_{p}^T \nabla\bfy_{p},\bfC_{p} ^{v}), \qquad \dot{\bfE}_{n+1} ^{v}=\mathbf{H}(\nabla\bfy_{p}^T \nabla\bfy_{p}, \nabla \varphi_{p},\bf E_{p}^v) \quad \forall \bfX \in \textrm{$\Omega_0$}, \label{eq:TimeDiscretization_Evol_eqn}
\end{align}
where $p=n$ results in an explicit time-discretization and $p=n+1$ results in an implicit one. We keep the discretization scheme general at this point and come back to it in Section~\ref{Sec:NumericalScheme}.
\subsubsection{Space Discretization}

Having discretized the equation in time, the next step is to discretize the equations in space. For this, we consider discrete partitions $\Omega_0^h = \bigcup_{e=1}^{\texttt{N}_e} \mathcal{E}^e$ of the continuous domain $\Omega_0$ into $\texttt{N}_e$ non-overlapping simplicial elements $\mathcal{E}^e$. The problem then is to look for approximate solutions $\bfy^h_{n+1}(\bfX)$ and $p_{n+1}^h(\bfX)$ in the finite dimensional subspaces of $\mathcal{Y}^h \subset \mathcal{Y}$ and $\mathcal{P}^h \subset \mathcal{P}$, and $\varphi^h_{n+1}$ in the finite dimensional subspace $\mathcal{V}^h \subset \mathcal{V}$. The finite element spaces for displacement and pressure in $\mathcal{Y}^h$ and $\mathcal{P}^h$ cannot be arbitrary but must be chosen such that the combination is \textit{inf-sup} stable (see, e.g., \cite{babuvska1973finite,brezzi1974existence}).
% \bsdelete{For the Abaqus UEL implementation, the spaces chosen is the three-dimensional analogue \cite{girault2012finite} of the conforming Crouzeix-Raviart space \cite{crouzeix1973conforming} and for} 
For our \texttt{FEniCSx} implementation we choose the lowest order Taylor-Hood \cite{taylor1973numerical} subspace as outlined in Section \ref{Sec:FEniCS-X}, which amounts to choosing piecewise quadratic approximations for the deformation field $\mathbf{y}^h_{n+1}(\bfX)$ and electric potential $\varphi^h_{n+1}(\bfX)$ and piecewise linear approximation for the pressure field $p^h_{n+1}(\bfX)$. It then follows that the approximate solutions $\bfy_{n+1}^h(\bfX)$, $\varphi_{n+1}^h(\bfX)$ and $p_{n+1}^h(\bfX)$ admit the forms
\begin{align}
    \bfy^h_{n+1}(\bfX) &= \underset{e=1}{\overset{\texttt{N}_e}{\mathbb{A}}} \left(\sum^{\texttt{N}_{m,e}}_{m=1}\mathbf{N}^{m,e}_{\bfy}(\bfX)\bfy^{e,m}_{n+1}\right) = \sum^{\texttt{N}_{m}}_{m=1}\mathbf{N}^{m}_{\bfy}(\bfX)\bfy^{m}_{n+1} \label{eq:Rep1}\\ 
    \varphi^h_{n+1}(\bfX) &= \underset{e=1}{\overset{\texttt{N}_e}{\mathbb{A}}} \left(\sum^{\texttt{N}_{m,e}}_{m=1}N^{m,e}_{\varphi}(\bfX)\varphi^{e,m}_{n+1}\right) = \sum^{\texttt{N}_{m}}_{m=1}N^{m}_{\varphi}(\bfX)\varphi^{m}_{n+1} \notag\\
    p^h_{n+1}(\bfX) &= \underset{e=1}{\overset{\texttt{N}_e}{\mathbb{A}}} \left(\sum^{\texttt{N}_{q,e}}_{q=1}N^{q,e}_{p}(\bfX)p^{e,q}_{n+1}\right) = \sum^{\texttt{N}_{q}}_{q=1}=N^{q}_{p}(\bfX)p^{q}_{n+1},\label{eq:Rep3}
\end{align}
where, $\mathbf{N}^{m,e}_{\bfy}(\bfX)$, $N^{m,e}_{\varphi}(\bfX)$ and $N^{q,e}_{p}(\bfX)$ are the element shape functions associated with the element degrees of freedom $\bfy_{n+1}^{e,m}$, $\varphi_{n+1}^{e,m}$ and $p_{n+1}^{e,q}$ respectively. Here, $\texttt{N}_{m,e}$ and $\texttt{N}_{q,e}$ are the total number of nodes in each element and the number of degrees of freedom per element for the approximation of $p_{n+1}^h$ respectively. After the assembly process using the assembly operator $\mathbb{A}(\cdot)$, $\mathbf{N}_{\bfy}^m(\bfX)$, $\mathbf{N}_{\bfy}^m(\bfX)$ and $\mathbf{N}_{\bfy}^m(\bfX)$ denote the global shape functions that result from the assembly process while, $\texttt{N}_{m}$ and $\texttt{N}_{q}$ are the total number of nodes in the partition $\Omega_0^h$ and the number of degrees of freedom for the approximation of $p_{n+1}^h$. We emphasize the previous remark that the shape functions $\mathbf{N}^{m,e}_{\bfy}(\bfX)$ and $N^{q,e}_{p}(\bfX)$ must be selected with care to ensure a stable formulation \cite{sussman1987finite}.

Given the spatial discretization, the weak form of the governing equations \eqref{BVP-F-hybrid-weak_disc} can be rewritten as 

\begin{align}
    \hspace*{-3ex}
    \left\{\begin{array}{ll}
    \displaystyle
     \underset{e=1}{\overset{\texttt{N}_e}{\mathbb{A}}}
     \int_{\mathcal{E}^e}
     \left[\bfS_{n+1}\left(\nabla\bfy^h_{n+1}, -\nabla\varphi^h_{n+1},\bfC^v_{n+1},\bfE^v_{n+1}\right) 
     + 
     p_{n+1}J_{n+1} \nabla \bfy_{n+1}^{-T}\right] \cdot \nabla \mathbf{v} \ \rm d \bfX
     =\\[10pt]
     \qquad \qquad \displaystyle
     \underset{e=1}{\overset{\texttt{N}_t}{\mathbb{A}}}
     \int_{\mathcal{E}^e}
     \mathbf{f}_{n+1}(\bfX)\cdot \mathbf{v} \ \rm d \bfX
     + 
     \underset{e=1}{\overset{\texttt{N}_t}{\mathbb{A}}}
     \int_{\partial \mathcal{E}^e}\overline{\textbf{t}}_{n+1}(\bfX)\cdot \mathbf{v} \ \rm d \bfX, \quad \forall \mathbf{v} \in \mathcal{Y}_0\\[10pt]
    \displaystyle
    \underset{e=1}{\overset{\texttt{N}_e}{\mathbb{A}}}
    \int_{\mathcal{E}^e}
    \left(\det\nabla\bfy_{n+1}(\bfX) 
    -\dfrac{1 + p_{n+1}/\kappa+\sqrt{4\sum_{r=1}^2\mu_r/\kappa + (1+p_{n+1}/\kappa)^2}}{2}\right)q \ \rm d \bfX = 0,\quad \forall \textrm{$q$}\in \mathcal{P}_0\\[10pt]
     \displaystyle
     \underset{e=1}{\overset{\texttt{N}_t}{\mathbb{A}}}
     \int_{\mathcal{E}^e}{\rm Div}\left[\bfD_{n+1}\left(\nabla\bfy_{n+1},-\nabla\varphi_{n+1},\bfC_{n+1}^v,\bfE_{n+1}^v\right)\right]\cdot \phi \ \rm d \bfX = \\[10pt]
     {\qquad \qquad \displaystyle\underset{e=1}{\overset{\texttt{N}_e}{\mathbb{A}}}\int_{\mathcal{E}^e}Q_{n+1}(\bfX)\phi \ \rm d \bfX 
     + 
     \displaystyle
     \underset{e=1}{\overset{\texttt{N}_{t_{\varphi}}}{\mathbb{A}}}
     \int_{\partial \mathcal{E}^e}\left(\sigma_{n+1}(\bfX)
     -
     \overline{\bfD}_{n+1}(\bfX)\cdot\bfN\right) \ \rm d \bfX,\quad \forall \textrm{$\phi$}\in \mathcal{V}_0}
     % \underbrace_{{\Pi^h}}
     \end{array}
     \right. . \label{eq:BVP-F-hybrid-galerkin-disc}
\end{align}
Furthermore, the integrals over each element $\mathcal{E}^e$ in \eqref{eq:BVP-F-hybrid-galerkin-disc} are calculated using a quadrature rule\footnote{Any quadrature rule of degree $K$ evaluates the integrands at quadrature points $\bfX_{\texttt{k}}$ along with corresponding weights $\omega_k$ where $\texttt{k}\in[2K-1]$. In practice, as is often done in standard FE implementation, the integrands from the physical element $\mathcal{E}^e$ are always mapped to a reference domain $\hat{\mathcal{E}}$ where the quadrature rules are specified.}, which results in a system of nonlinear algebraic equations for the nodal degrees-of-freedom $\bfy^m_{n+1}$, $\varphi^m_{n+1}$ and $p^q_{n+1}$ that need to be solved in conjunction with another system of nonlinear algebraic equations for the internal variables $\bfC^{v^h}_{n+1}(\bfX)$ and $\bfE^{v^h}_{n+1}(\bfX)$ at each quadrature point.
% \bsdelete{
% The representations in \eqref{eq:Rep1}-\eqref{eq:Rep3}, together with the time-discretized weak form \eqref{BVP-F-hybrid-weak}, lead to a system of coupled nonlinear algebraic equations for the variables $\bfy_{n+1}^m$, $\varphi_{n+1}^m$, and $p_{n+1}^q$. These equations are dependent on the internal variables $\bfC^{v^h}_{n+1}$ and $\bfE^{v^h}_n$ at the quadrature integration points used to evaluate the integral in (\ref{BVP-F-hybrid-weak}) and} 
With slight abuse of notation, this can be written as 
\begin{align}
    \mathcal{F}(\bfy^m_{n+1},\varphi^m_{n+1},p^q_{n+1},\bfC^{v^h}_{n+1},\bfE^{v^h}_{n+1}) = 0.\label{eq:Nonlin_Algebraic_1}
\end{align}
for the nodal degrees of freedom $\bfy^m_{n+1}$, $\varphi^m_{n+1}$ and $p^q_{n+1}$ together with
% \bsdelete{
% %
% Similarly, the coupled systems of nonlinear algebraic equations resulting from (\ref{eq:TimeDiscretization_Evol_eqn}) for the internal variables $\bfC^{v^h}_n$ and $\bfE^{v^h}_n$ at the quadrature points can be represented as: }
%
\begin{align}
    \mathcal{G}(\bfy^h_{n+1},\bfC^{v^h}_{n+1},\dot{\bfC}^{v^h}_p) = 0 \qquad \textrm{and} \qquad \mathcal{H}(\bfy^h_{n+1},\varphi^h_{n+1},\bfE^{v^h}_{n+1},\dot{\bfE}^{v^h}_p) = 0.\label{eq:Nonlin_Algebraic_2}
\end{align}
for the internal variable at each quadrature point. These are often solver using a staggered-scheme by employing a nonlinear solver at each quadrature point for the internal variables embedded inside a global nonlinear solver for the nodal degrees of freedom. The solution procedure is iterated until convergence is reached for both the solvers. We defer the details of the solver along with the algorithm to Section~\ref{Sec:NumericalScheme}.
%

% \bsdelete{A staggered solver that consists of an explicit time integration for solving \eqref{eq:Nonlin_Algebraic_2} and a Newton-like method for solving (\ref{eq:Nonlin_Algebraic_1}) is illustrated in \ref{Appendix:Staggered_Solver}}
% The finite dimensional subspaces are:
% %
% \begin{align*}
%     \mathcal{Y}^h &= {\bfy^h \in \left[C^0(\bfX)\right]^3 \cap \mathcal{Y}: \bfy^{h,e}_n = \sum^{\texttt{N}_m}_{m=1}\mathbf{N}^{m}_{\star}(\bfX)\bfy^{e,m}_n, \quad \forall e=1,...,\texttt{N}_e\}\\
%     \mathcal{V}^h &= {\varphi^h \in \left[C^0(\bfX)\right] \cap \mathcal{V}: \varphi^{h,e}_n = \sum^{\texttt{N}_m}_{m=1}\mathbf{N}^{m}_{\star}(\bfX)\varphi^{e,m}_n, \quad \forall e=1,...,\texttt{N}_e\}\\
%     \mathcal{P}^h &= {p^h \in \mathcal{P}: p^{h,e}_n = \sum^{3}_{k=0}\mathbf{N}^{p,k}_{\star}(\bfX)p^{e,k}_n, \quad \forall e=1,...,\texttt{N}_e\},
% \end{align*}
% %
% where, $C^0(\bfX)$ denotes the set of continuous functions on $\bfX$, $\bfy_n^{e,m}$ and $\varphi_n^{e,m}$ are the displacements and potentials at the node $m$ in element $e$ at time $t_n$, while $p_n^{e,k}$ denote the value of the pressure field and its gradients 

\section{Numerical Implementation}\label{Sec:NumericalScheme}
Having outlined the governing equations along with their discretizations in Section~\ref{Sec:Problem}, we turn our attention to their solution in this section. Specifically we implement various numerical schemes to study the material response of an acrylate elastomer VHB 4910 under different loading conditions. The material parameters for VHB 4910 determined from the experiments in \cite{hossain2012experimental} and \cite{qiang2012experimental} were reported in \cite{ghosh2021two} and are also included in \ref{Sec:Material_Parameters} for completeness. First, the governing equations are solved for the special case of homogeneous applied deformation and electric fields, which reduces the equations to one dimensional nonlinear equations. These one dimensional equations still yield profound insights that are useful in designing algorithms for the full-field three dimensional equations presented in Section~\ref{Sec:Problem}. We investigate three time-discretization schemes. We vary the order (1\textsuperscript{st} order vs 5\textsuperscript{th} order) and nature (explicit vs implicit) of the schemes and their overall effect on the time-to-solution (TTS). Secondly, we then choose a given scheme to spell out the full solution algorithm to solve \eqref{BVP-F-hybrid-weak_disc} in the FE framework \texttt{FEniCSx}.

\subsection{1-D Problem: Uniaxial Tension \& Electric Field}\label{subsec:1D_problem_Numerics}
\begin{figure}[h!]
    \centering
    \begin{subfigure}[H]{0.45\textwidth}
    \centering
        \includegraphics[width=\linewidth]{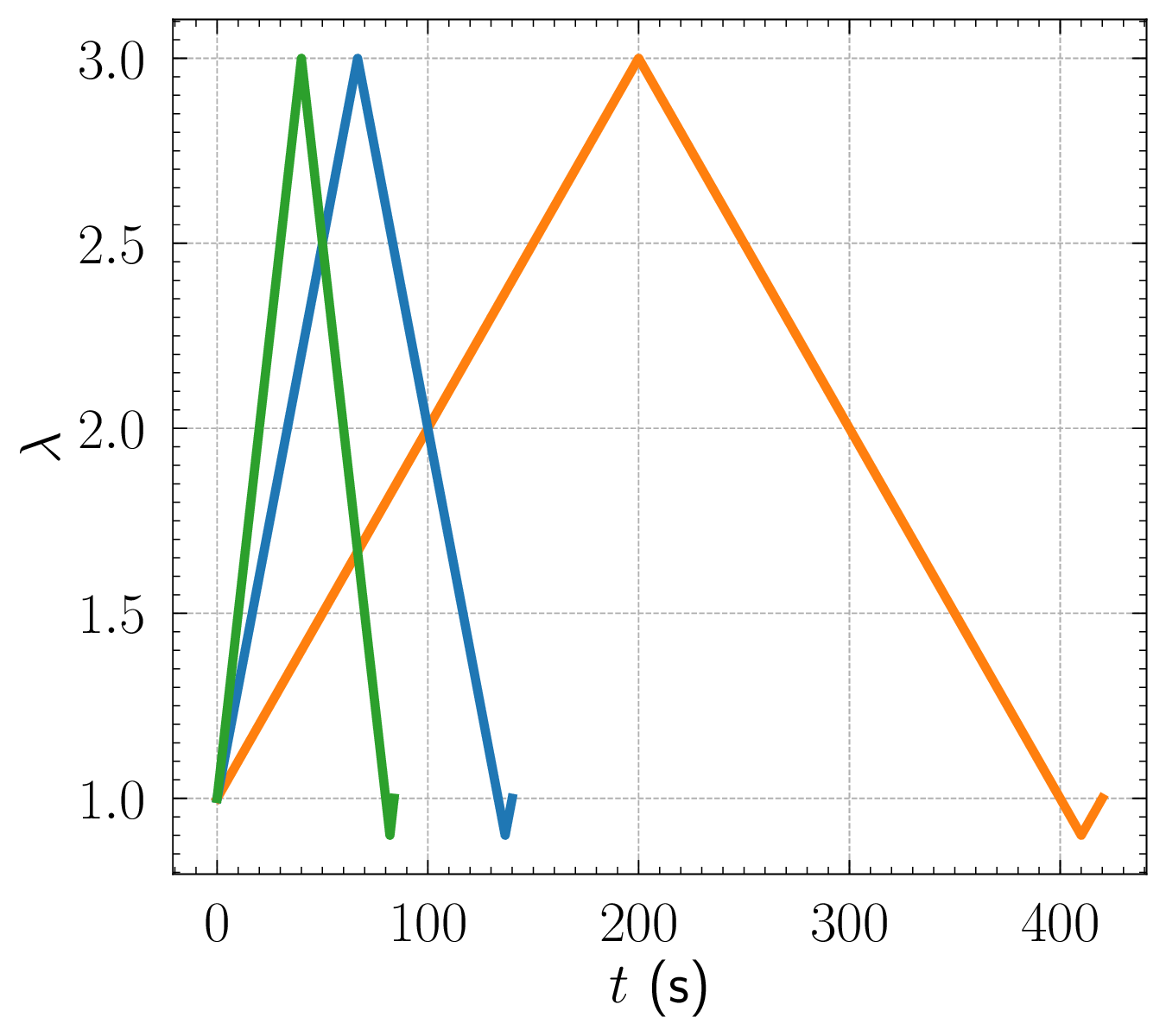}
        % \caption{}
    \end{subfigure}\hfill\begin{subfigure}[H]{0.45\textwidth}
    \centering
        \includegraphics[width=\linewidth]{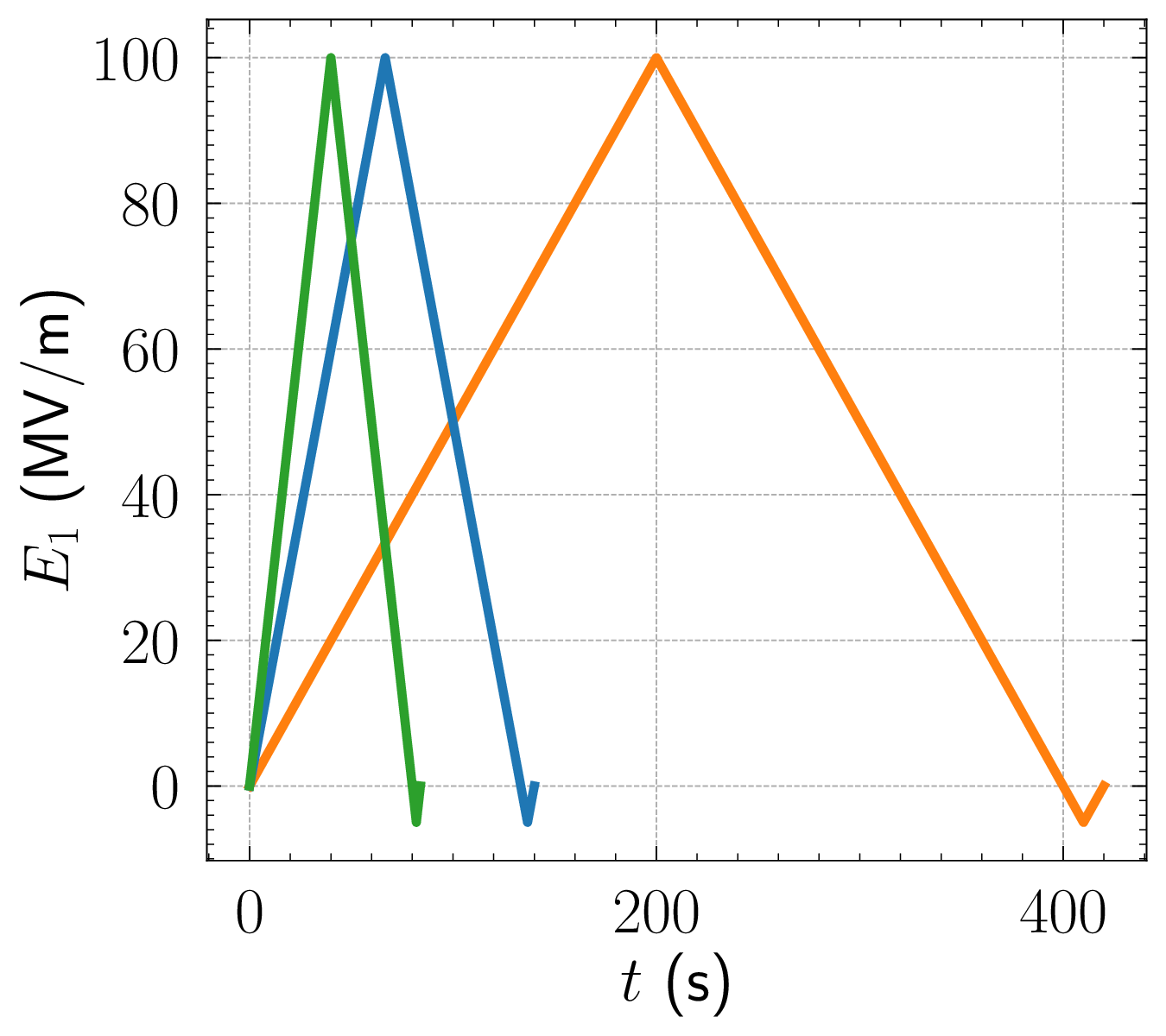}
        % \caption{}
    \end{subfigure}\\
    \begin{subfigure}[H]{0.75\textwidth}
    \centering
        \includegraphics[width=\linewidth]{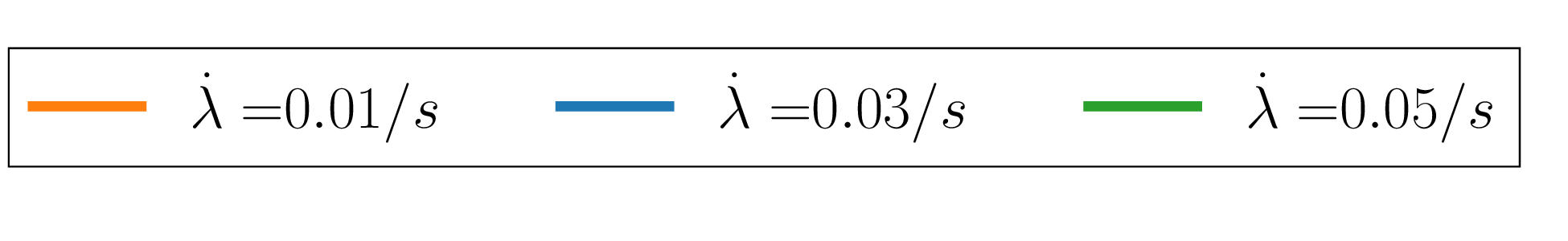}
        % \caption{}
    \end{subfigure}
    \caption{Applied Stretch $\lambda(t)$ and lagrangian electric field $E_1(t)$ considered for solving the system of nonlinear ODEs \eqref{eq:ODES_Cv_Ev} in Sections (\ref{sec:RK5_ODE})-(\ref{sec:BDF}). The loading rates are chosen due to their practical relevance in experiments (see \cite{hossain2015comprehensive})}
    \label{fig:applied_Stretch_ElecField_triangle}
\end{figure}

Here, we spell out the material response of VHB 4910 under some canonical loadings. Specifically, we assume homogeneous uniaxial deformations and homogeneous uniaxial electric fields, namely 
\begin{equation}
    \bfF(t) = \begin{bmatrix}
        \lambda(t) &0 &0 \\
        0 &\dfrac{1}{\sqrt{\lambda(t)}} &0\\
        0 & 0 & \dfrac{1}{\sqrt{\lambda(t)}}
    \end{bmatrix}, \qquad
    \bfE(t) = \begin{bmatrix}
        E_1(t) \\ 0 \\ 0
    \end{bmatrix},\label{eq:appliedDeformations_elecF}
\end{equation}
where $\lambda(t)$ and $E_1(t)$ are the applied stretch and Lagrangian electric field respectively. The uniaxial nature of $\bfF(t)$ and $\bfE(t)$ result in the internal variables also taking a special form, namely
\begin{align}
    \bfC^v(t) = \begin{bmatrix}
        \lambda^v(t) &0 &0 \\
        0 &\dfrac{1}{\sqrt{\lambda^v(t)}}(t) &0\\
        0 & 0 & &\dfrac{1}{\sqrt{\lambda^v(t)}}(t)
    \end{bmatrix}, \qquad
    \bfE(t) = \begin{bmatrix}
        E_1(t) \\ 0 \\ 0
    \end{bmatrix},\label{eq:internal_variables}
\end{align}
Now, for the above canonical loadings, the system of IBVPs \eqref{BVP-F}-\eqref{Evol-Eq-Ev2} results in a stress field of the form 
\begin{align}
    \bfS(t) = \begin{bmatrix}
        S_{11}(t) &0 &0 \\
        0 &0 &0\\
        0 & 0 &0
    \end{bmatrix},
\end{align}
where, $S_{11}(t)$ can be derived from (\ref{Proposed CR- P-K stress}) and the hydrostatic pressure $p$ appearing in (\ref{Proposed CR- P-K stress}) can be obtained by equating $S_{22}(t)=S_{33}(t)=0$. The evolution equations reduce to a system of two nonlinear ordinary differential equations for the internal variables $C^v$ and $E^v$, namely
\begin{align}
    \begin{aligned}
        \dot{\bfC}^v &= \mathbf{G}(\bfF^T\bfF, \bfC^v)\equiv\hat{G}(\lambda, \lambda^v),\\
        \dot{\bfE}^v &= \mathbf{H}(\bfF^T\bfF, \bfE, \bfE^v)\equiv\hat{H}(\lambda, E_1, E^v)
    \end{aligned}
    \label{eq:ODES_Cv_Ev}
\end{align}
where $\hat{G}$ and $\hat{H}$ are derived from \eqref{Evol-Eq-Cv2} and \eqref{Evol-Eq-Ev2} which needs to be solved via numerical time-integration schemes.
In order to illustrate the differences between different time-integration algorithms, we consider three different algorithms: \emph{(i)} $5^{th}$ order Runge-Kutta method that has previously been demonstrated in ~\cite{ghosh2021two}, \emph{(ii)} backward Euler or first order implicit Euler algorithm that is often used in practice due to its stability for linear ODEs, and \emph{(iii)} a high-order implicit method \cite{byrne1975polyalgorithm} based on the backward differentiation formulas (BDF) that provides adaptive switching of the integration order based on the stiffness of the ODE. For each of the above algorithms we solve the ODEs by for the following loadings $E_t(t)$ and $\lambda(t)$:
\begin{align}
    E_t(t) = \begin{cases}
        \dot{E}_1 t & 0 \leq t < t_1\\
        \bar{E}_{\max} - \dot{E}_1 t & t_1 \leq t < t_2\\
         \bar{E}_{\min} + \dot{E}_1 t & t_2\leq t < T_f
    \end{cases},\qquad 
    \lambda(t) = \begin{cases}
         1 + \dot{\lambda}_1 t & 0 \leq t < t_1\\
         \bar{\lambda}_{\max} - \dot{\lambda}_1 t & t_1 \leq t < t_2\\
         \bar{\lambda}_{\min} + \dot{\lambda}_1 t & t_2\leq t < T_f 
    \end{cases},
\end{align}
which correspond to a single period of a triangular wave (see Fig.~\ref{fig:applied_Stretch_ElecField_triangle}). Next, we discretize the system of nonlinear ODEs \eqref{eq:ODES_Cv_Ev} at $N+1$ steps $\{t_n\}_{n=0}^N$ with $t_0$ corresponding to the initial condition ($t_0=0$ in this case) and solve the equations at each time $t_{n+1}$, $n\in [1, N-1]$.
% \begin{figure}[h!]
%     \centering
%     \includegraphics[width=0.75\linewidth]{FiguresNew/electric_mechanical.eps}
%     \caption{Stress-stretch response of VHB 4910~\cite{ghosh2021two,hossain2015comprehensive} subjected to uniaxial loading unloading. Both the coupled electro-mechanical and mechanical responses are plotted to show the differences between them}
%     \label{fig:applied_Stretch_ElecField_triangle}
% \end{figure}

\begin{figure}[h!]
    \centering
    \begin{subfigure}[H]{0.33\linewidth}
        \centering
        \includegraphics[width=\linewidth]{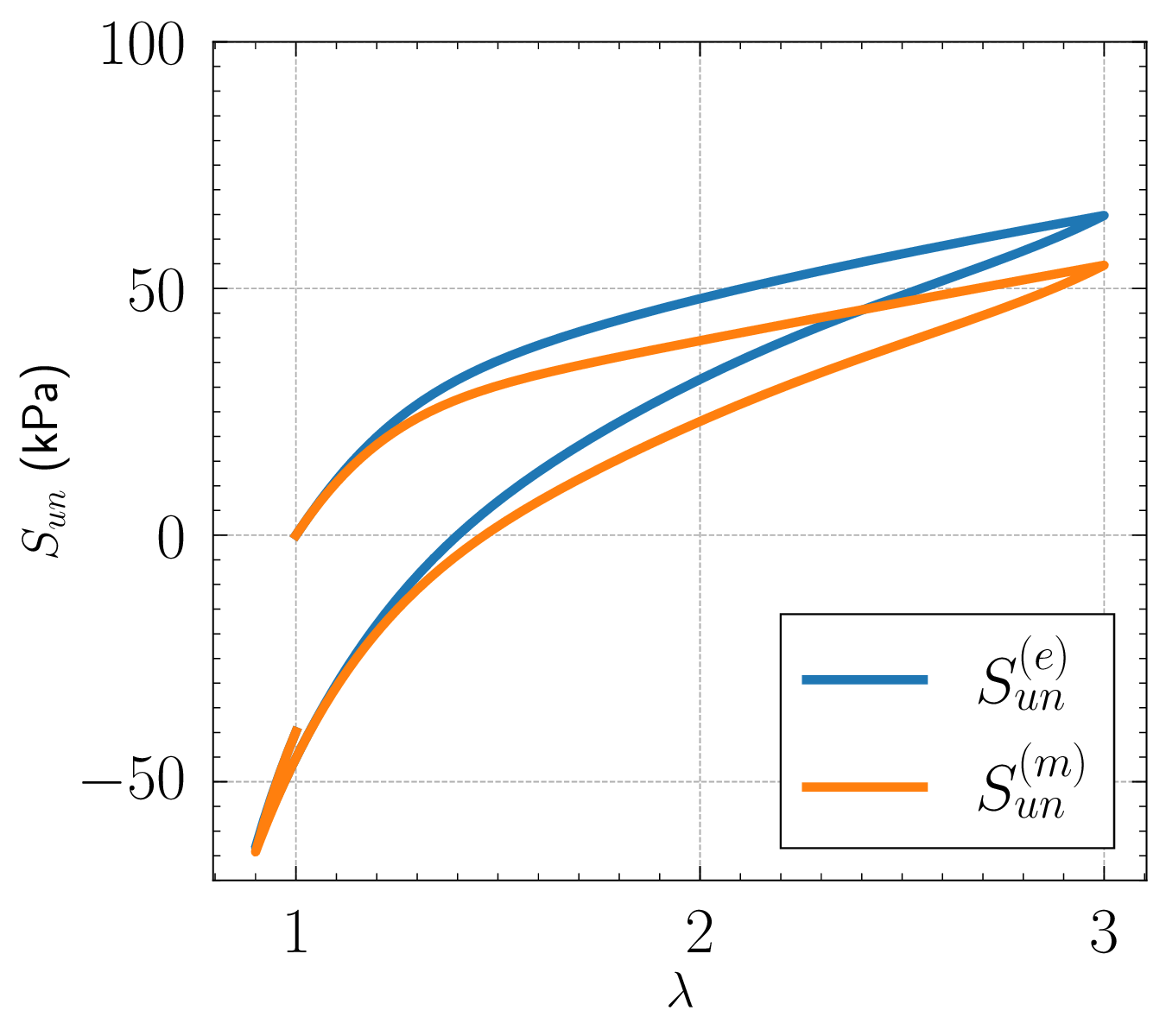}
        \caption{$\dot{\lambda} = 0.01/s$}
    \end{subfigure}\begin{subfigure}[H]{0.33\textwidth}
        \centering
        \includegraphics[width=\linewidth]{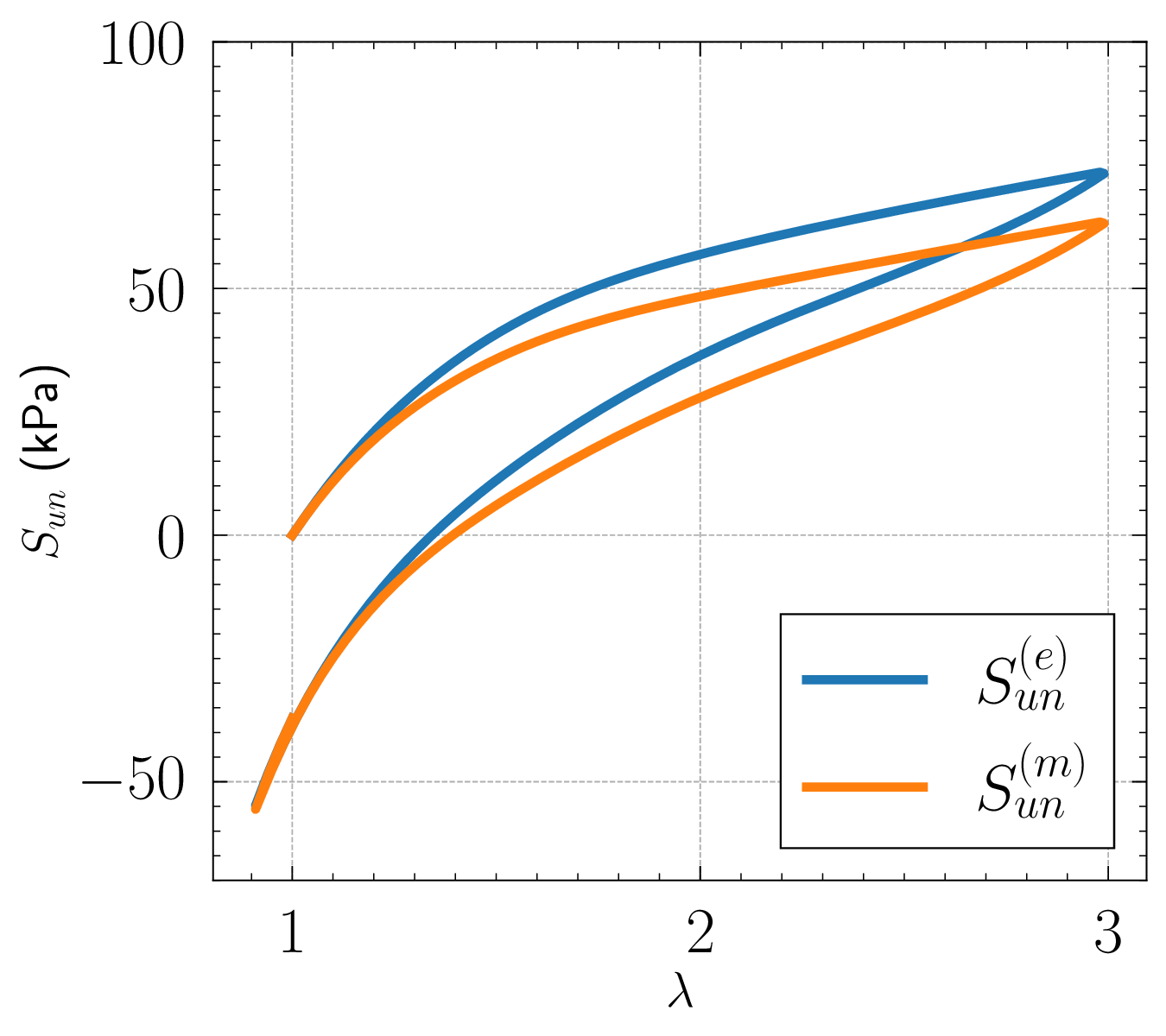}
        \caption{$\dot{\lambda} = 0.03/s$}
    \end{subfigure}\begin{subfigure}[H]{0.33\textwidth}
        \centering
        \includegraphics[width=\linewidth]{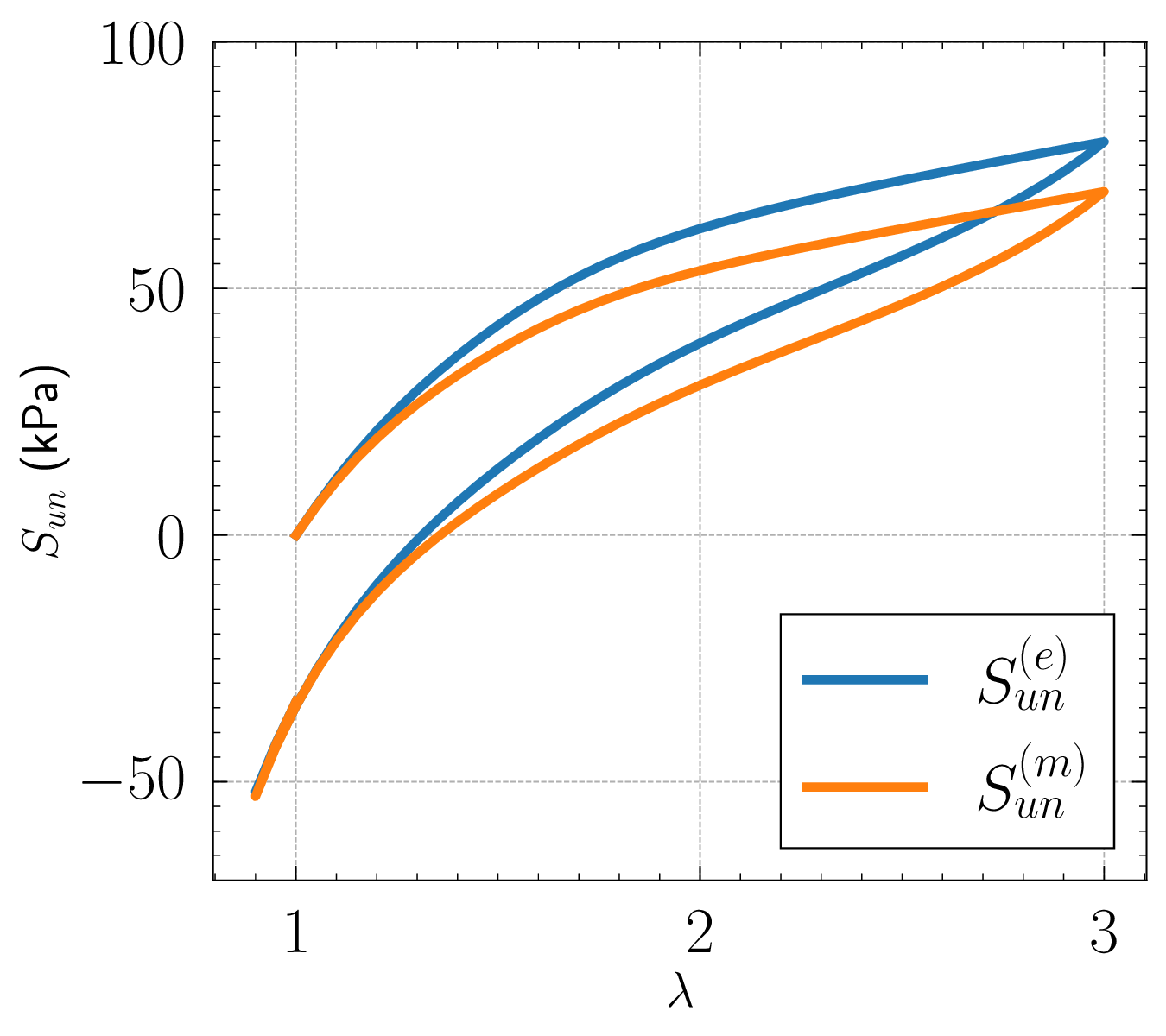}
        \caption{$\dot{\lambda} = 0.05/s$}
    \end{subfigure}
    \caption{The stress-stretch response of VHB 4910 subjected to electro-mechanical ($S^{(e)}_{un}$) and purely-mechanical ($S^{(m)}_{un}$) loadings for three different loading rates. In the electro-mechanical case, we consider both the electric and mechanical dissipation whereas in the purely-mechanical case we consider the effect of only mechanical dissipation}\label{fig:stress_stretch_elec_mech}
\end{figure}

\begin{figure}[h!]
    \centering
    \begin{subfigure}[H]{0.33\linewidth}
        \centering
        \includegraphics[width=\linewidth]{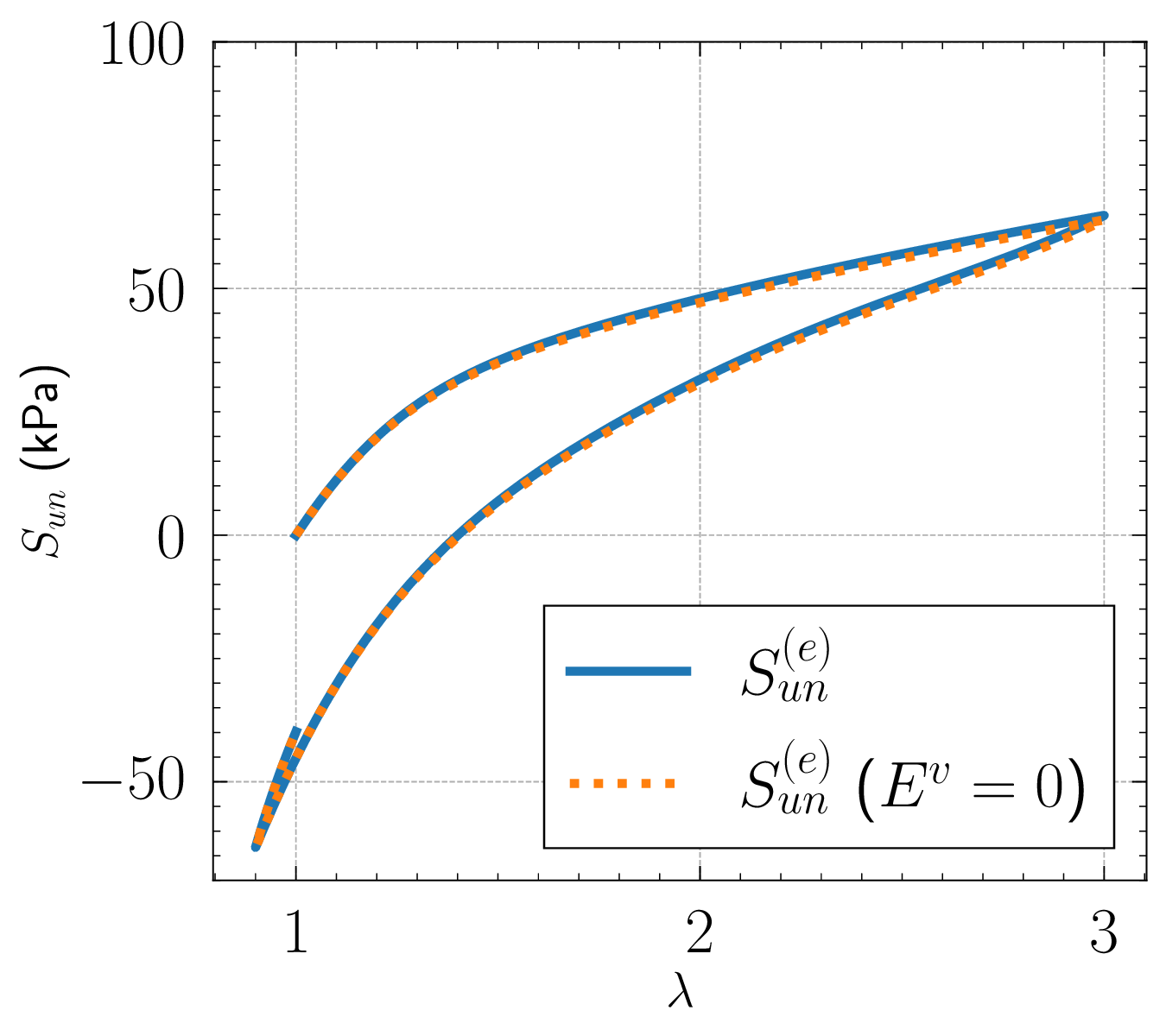}
        \caption{$\dot{\lambda} = 0.01/s$}
    \end{subfigure}\begin{subfigure}[H]{0.33\textwidth}
        \centering
        \includegraphics[width=\linewidth]{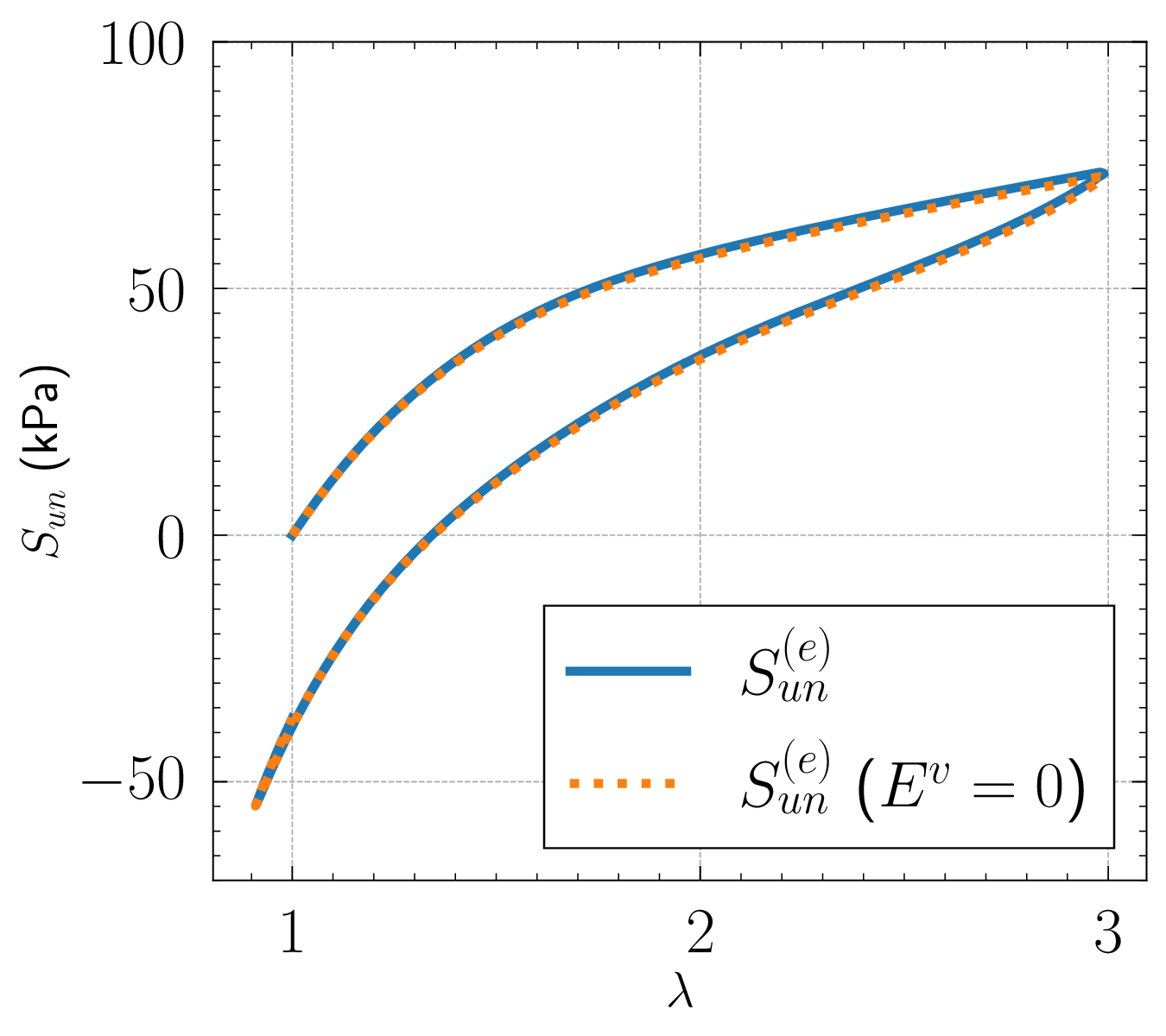}
        \caption{$\dot{\lambda} = 0.03/s$}
    \end{subfigure}\begin{subfigure}[H]{0.33\textwidth}
        \centering
        \includegraphics[width=\linewidth]{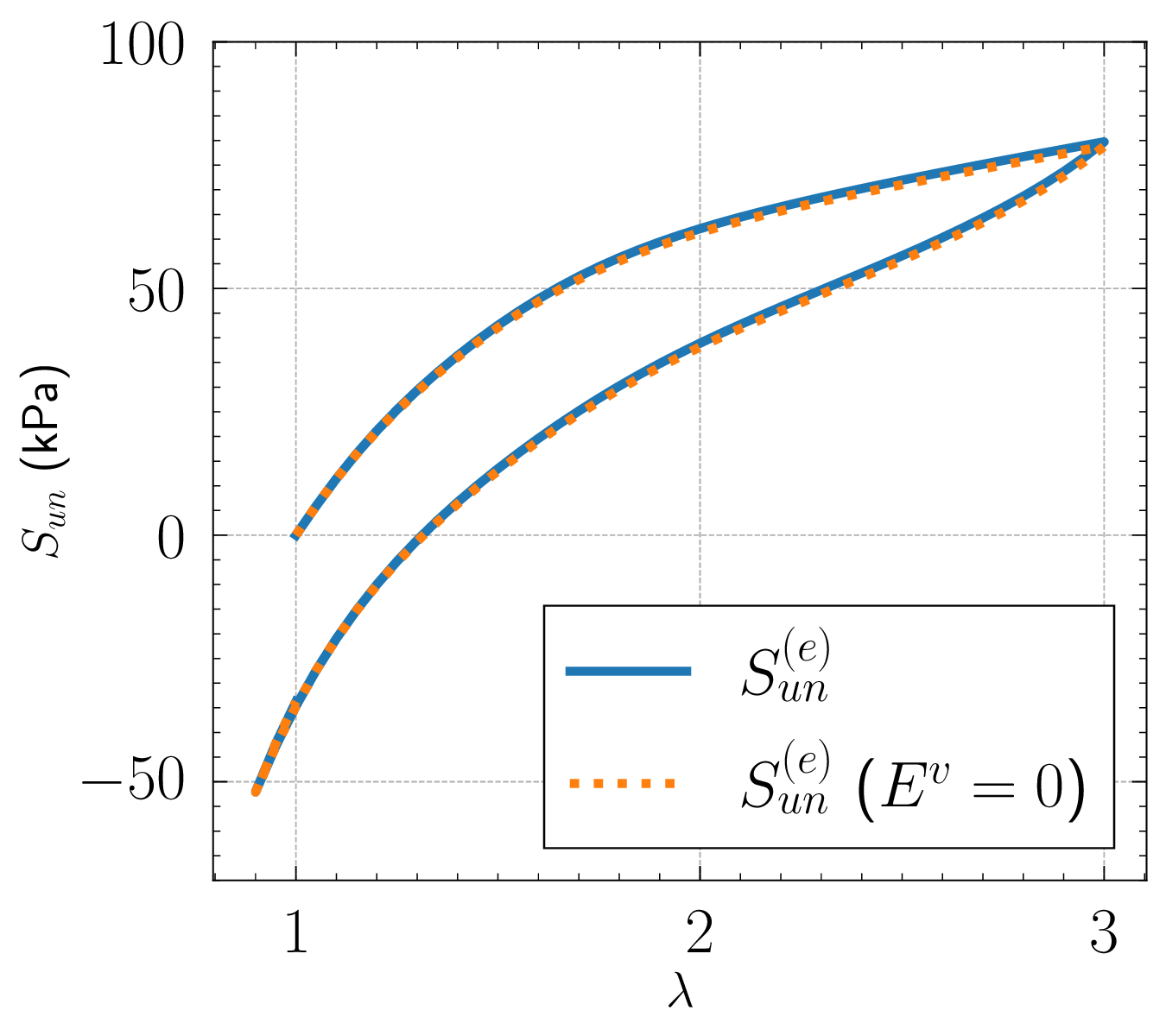}
        \caption{$\dot{\lambda} = 0.05/s$}
    \end{subfigure}
    \caption{The stress-stretch response of VHB 4910 subjected to electro-mechanical ($S^{(e)}_{un}$) with and without electric dissipation ($E^v$). We note that the stresses are practically indistinguishable from each other and hence for monotonic loadings explicit modeling of the evolution of $E^v$ may not be necessary.}
    \label{fig:response_Ev_vs_noEv}
\end{figure}

\subsubsection{Using a $5$\textsuperscript{th}-order Runge-Kutta solver}\label{sec:RK5_ODE}
In this section we present a $5^{th}$ order Runge-Kutta method (RK5) to solve the system of nonlinear ODEs \eqref{eq:ODES_Cv_Ev} (see, e.g. \cite{ghosh2021two} as well as \cite{ghosh2021nonlinear,shrimali2023delayed,shrimali2023nonlinear} for the analogous case of pure mechanical dissipation). The discretized equations at each time-step ($t_n$) are given by
\begin{align}
    {\bfC}_{t=t_{n+1}}^{v}={\bfC}_{t=t_n}^{v}+\frac{\Delta t}{90}\left(7 \mathbf{k}_1+32 \mathbf{k}_3+12 \mathbf{k}_4+32 \mathbf{k}_5+7 \mathbf{k}_6\right),
\end{align}
where
\begin{align}
    \begin{aligned}
        & \mathbf{k}_1=\mathbf{G}\left(t_n, {\bfC}_{t_n}^{v}\right) \\
        & \mathbf{k}_2=\mathbf{G}\left(t_n+\Delta t / 2, {\bfC}_{t_n}^{v}+\mathbf{k}_1 \Delta t / 2\right) \\
        & \mathbf{k}_3=\mathbf{G}\left(t_n+\Delta t / 4, {\bfC}_{t_n}^{v}+\left(3 \mathbf{k}_1+\mathbf{k}_2\right) \Delta t / 16\right) \\
        & \mathbf{k}_4=\mathbf{G}\left(t_n+\Delta t / 2, {\bfC}_{t_n}^{v}+\mathbf{k}_3 \Delta t / 2\right) \\
        & \mathbf{k}_5=\mathbf{G}\left(t_n+3 \Delta t / 4, {\bfC}_{t_n}^{v}+3\left(-\mathbf{k}_2+2 \mathbf{k}_3+3 \mathbf{k}_4\right) \Delta t / 16\right) \\
        & \mathbf{k}_6=\mathbf{G}\left(t_n+\Delta t, {\bfC}_{t_n}^{v}+\left(\mathbf{k}_1+4 \mathbf{k}_2+6 \mathbf{k}_3-12 \mathbf{k}_4+8 \mathbf{k}_5\right) \Delta t / 7\right)
    \end{aligned}\label{eq:RK5_Cv}
\end{align}
for the internal variable $\bfC^v$. Similarly, for the internal variable $\bfE^v$, we have
\begin{align}
    \bfE_{t=t_{n+1}}^{v}={\bfE}_{t=t_n}^{v}+\frac{\Delta t}{90}\left(7 \mathbf{l}_1+32 \mathbf{l}_3+12 \mathbf{l}_4+32 \mathbf{l}_5+7 \mathbf{l}_6\right)
\end{align}
with
\begin{align}
    \begin{aligned}
        & \mathbf{l}_1=\bfH\left(t_n, {E}_{t_n}^{v}\right) \\
        & \mathbf{l}_2=\bfH\left(t_n+\Delta t / 2, {E}_{t_n}^{v}+\mathbf{l}_1 \Delta t / 2\right) \\
        & \mathbf{l}_3=\bfH\left(t_n+\Delta t / 4, {E}_{t_n}^{v}+\left(3 \mathbf{l}_1+\mathbf{l}_2\right) \Delta t / 16\right) \\
        & \mathbf{l}_4=\bfH\left(t_n+\Delta t / 2, {E}_{t_n}^{v}+\mathbf{l}_3 \Delta t / 2\right) \\
        & \mathbf{l}_5=\bfH\left(t_n+3 \Delta t / 4, {E}_{t_n}^{v}+3\left(-\mathbf{l}_2+2 \mathbf{l}_3+3 \mathbf{l}_4\right) \Delta t / 16\right) \\
        & \mathbf{l}_6=\bfH\left(t_n+\Delta t, {E}_{t_n}^{v}+\left(\mathbf{l}_1+4 \mathbf{l}_2+6 \mathbf{l}_3-12 \mathbf{l}_4+8 \mathbf{l}_5\right) \Delta t / 7\right),
    \end{aligned}\label{eq:RK5_Ev}
\end{align}
for $n\in [1, N-1]$.
\begin{remark}[$5$\textsuperscript{th} order accuracy]
    RK5, by definition, is $5$\textsuperscript{th} order accurate in time. This gives better accuracy but also imposes a severe constraint on the time-step size ($\Delta t$) particularly to step-through \eqref{eq:RK5_Ev}.
\end{remark}
We solve \eqref{eq:ODES_Cv_Ev} using RK5 and the report the results in Fig.~\ref{fig:stress_stretch_method_compare} for three different electromechanical loadings.
\subsubsection{Using Backward Euler (BE) for the 1-D problem}\label{sec:BE_ODE}
In this section we present the Backward Euler algorithm (BE) to solve \eqref{eq:ODES_Cv_Ev}. The discrete counterpart of \eqref{eq:ODES_Cv_Ev} for a BE step ($t_{n+1}$) are given by 
\begin{align}
    \begin{aligned}
        C^v_{t=t_{n+1}} &= G(\lambda_{t=t_{n+1}}, C^{v}_{t=t_{n+1}}),\\
        E^v_{t=t_{n+1}} &= H(\lambda_{t=t_{n+1}}, E_{t=t_{n+1}}, E^{v}_{t=t_{n+1}}),
    \end{aligned}
\end{align}
which in turn are nothing but two nonlinear algebraic equations for $C^v_{n+1}$ and $E^v_{n+1}$ respectively. We implement the algorithm in \texttt{python} using the scientific python ecosystem (\texttt{NumPy} for array operations and \texttt{Numba} to just-in-time (\texttt{JIT}) compile performance intensive code wherever possible). We solve the discretized equations using the nonlinear solvers in \texttt{scipy} \cite{scipy}. The results obtained by applying BE to solve \eqref{eq:ODES_Cv_Ev} are also summarized in Fig.~\ref{fig:stress_stretch_method_compare}.
\subsubsection{Using BDF}\label{sec:BDF}
In this section we present an adaptive implicit time-integration algorithm for solving \eqref{eq:ODES_Cv_Ev} based on the backward differentiation formulas (BDF) (see \cite{byrne1975polyalgorithm}) of varying order. Note that BE is a special case of BDF when order=1. In this section we present results for the same uniaxial loadings considered in sections~\ref{sec:BE_ODE} and \ref{sec:RK5_ODE} with BDF with order set to 2. The discrete counterparts of \eqref{eq:ODES_Cv_Ev} read as
\begin{align}
    {C^v}_{n+2}-\frac{4}{3} {C^v}_{n+1}+\frac{1}{3} {C^v}_n=\frac{2}{3} \Delta t f\left(t_{n+2}, {C^v}_{n+2}\right),
\end{align}
and
\begin{align}
    {E^v}_{n+2}-\frac{4}{3} {E^v}_{n+1}+\frac{1}{3} {E^v}_n=\frac{2}{3} \Delta t f\left(t_{n+2}, {E^v}_{n+2}\right)
\end{align}
respectively.
The \texttt{python} wrappers for the Fortran solvers (\url{https://www.netlib.org/ode/vode.f}) based on BDF \cite{brown1989vode} are available through \texttt{scipy.integrate.ode} interface. Note that we report results for BDF (order=2) in this section.
\begin{figure}[h!]
    \centering
    \begin{subfigure}[H]{0.33\linewidth}
        \centering
        \includegraphics[width=\linewidth]{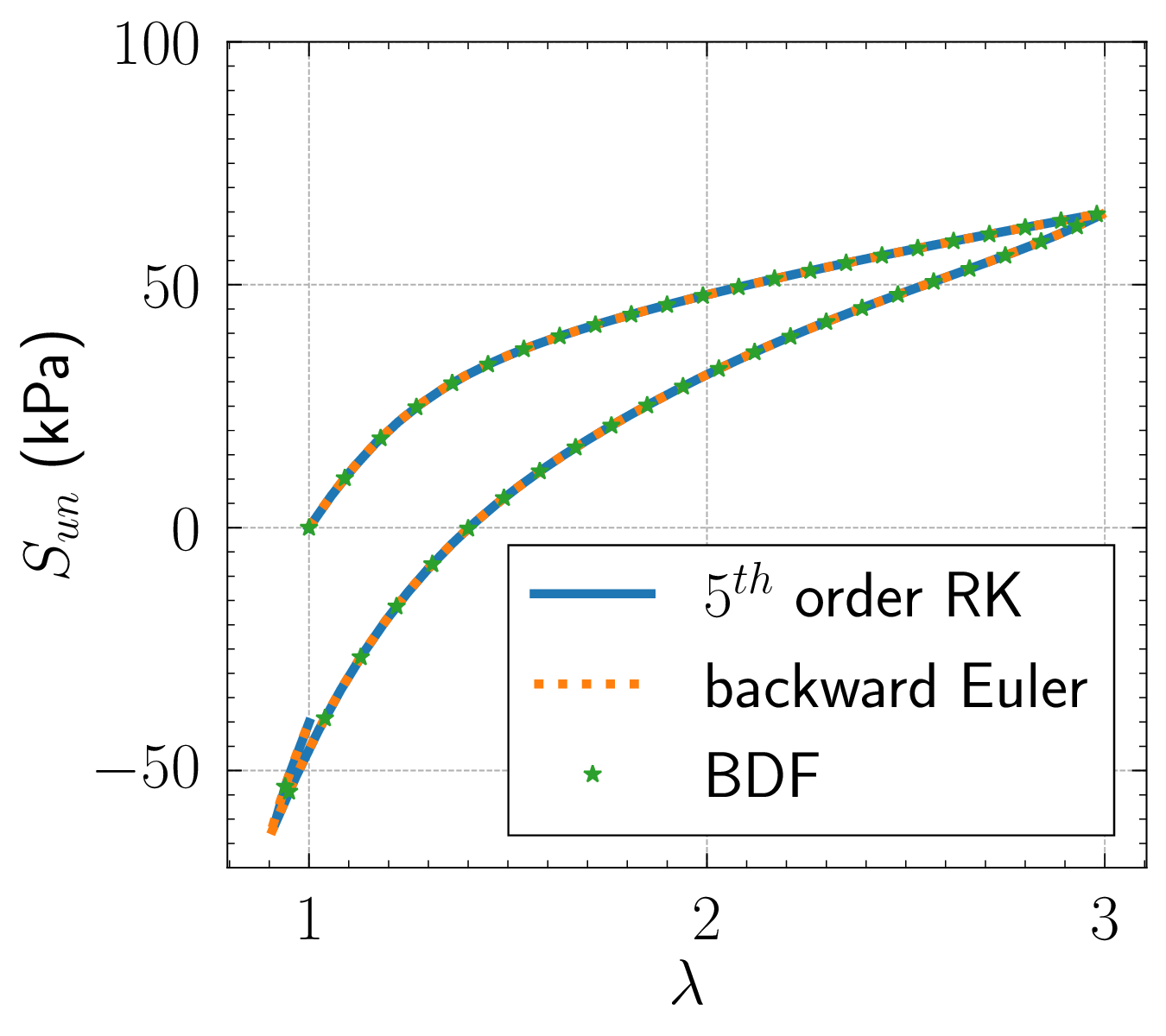}
        \caption{$\dot{\lambda} = 0.01/s$}
    \end{subfigure}\begin{subfigure}[H]{0.33\textwidth}
        \centering
        \includegraphics[width=\linewidth]{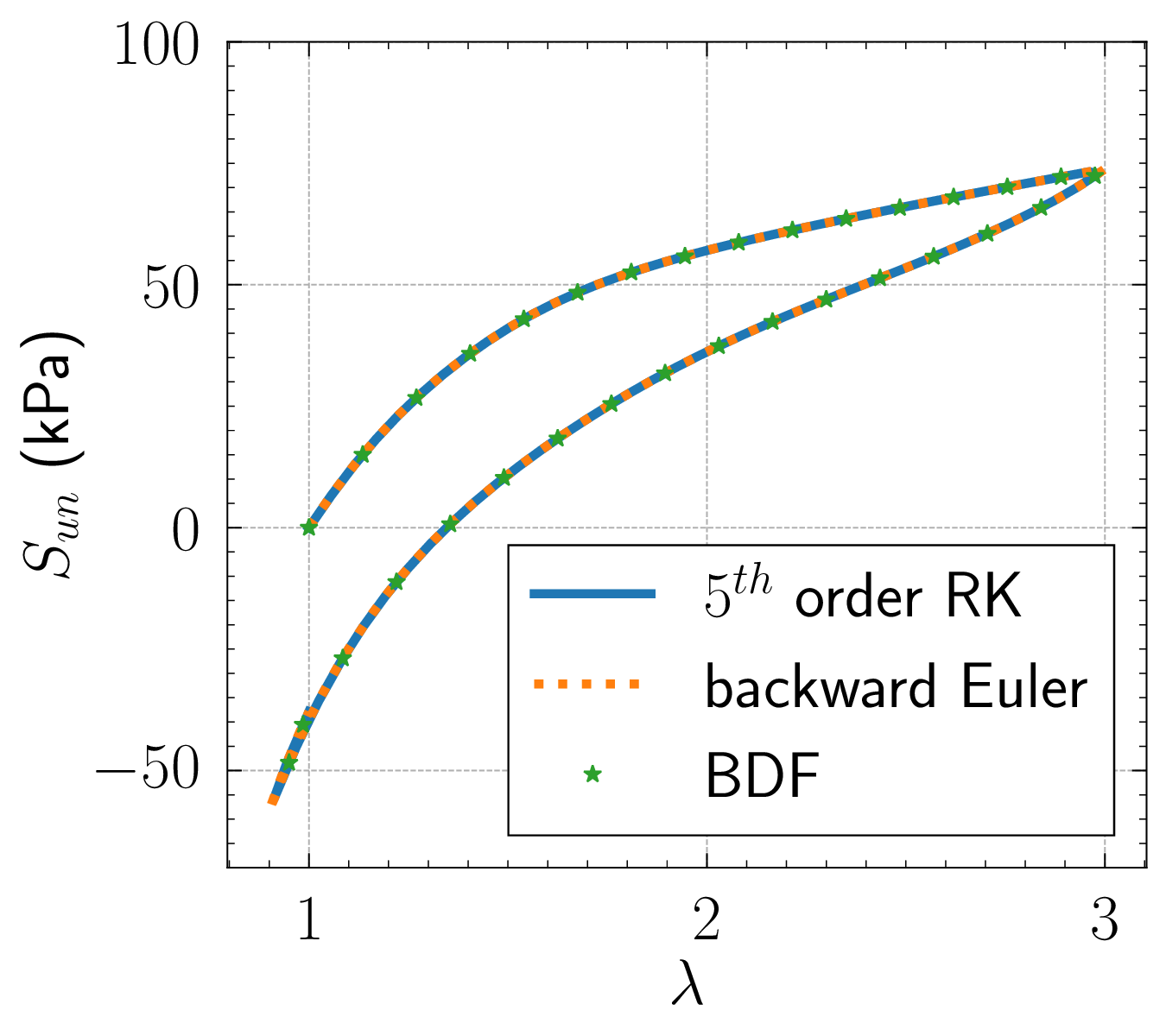}
        \caption{$\dot{\lambda} = 0.03/s$}
    \end{subfigure}\begin{subfigure}[H]{0.33\textwidth}
        \centering
        \includegraphics[width=\linewidth]{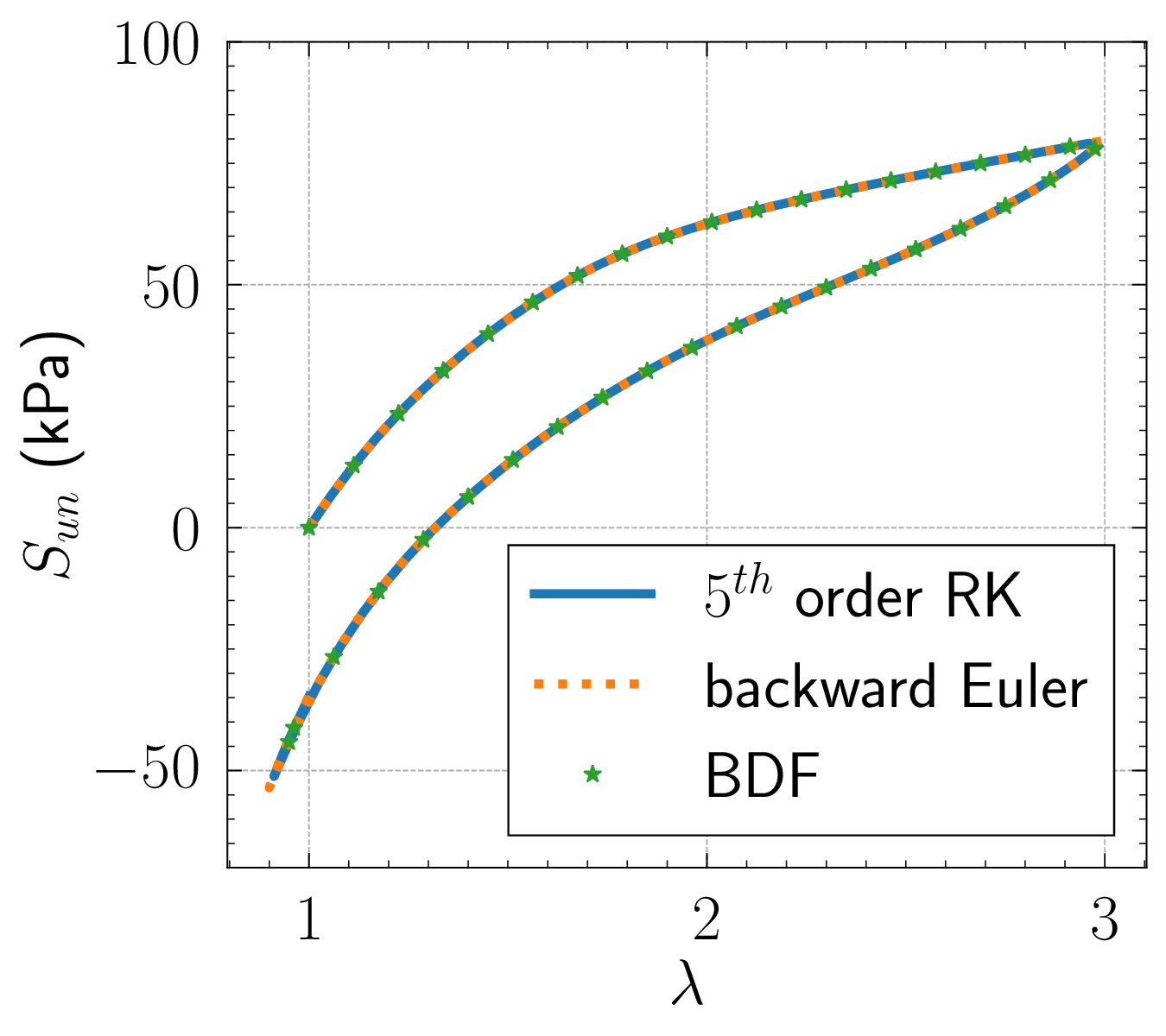}
        \caption{$\dot{\lambda} = 0.05/s$}
    \end{subfigure}
    \caption{The stress-stretch response of VHB 4910 subject to the loadings \eqref{eq:appliedDeformations_elecF} computed using the different time-integration algorithms summarized in Sections~(\ref{sec:RK5_ODE})-(\ref{sec:BDF}). Note that the maximum stable time increment for 
 the RK5 solver was $10^{-6}$ seconds, whereas for BE it was $10^{-2}$ seconds. The maximum stable time increment for BDF was in the range ($10^{-4}$, $10^{-3}$) seconds.}\label{fig:stress_stretch_method_compare}
\end{figure}
\begin{figure}[h!]
    \centering
    \begin{subfigure}[H]{0.33\linewidth}
        \centering
        \includegraphics[width=\linewidth]{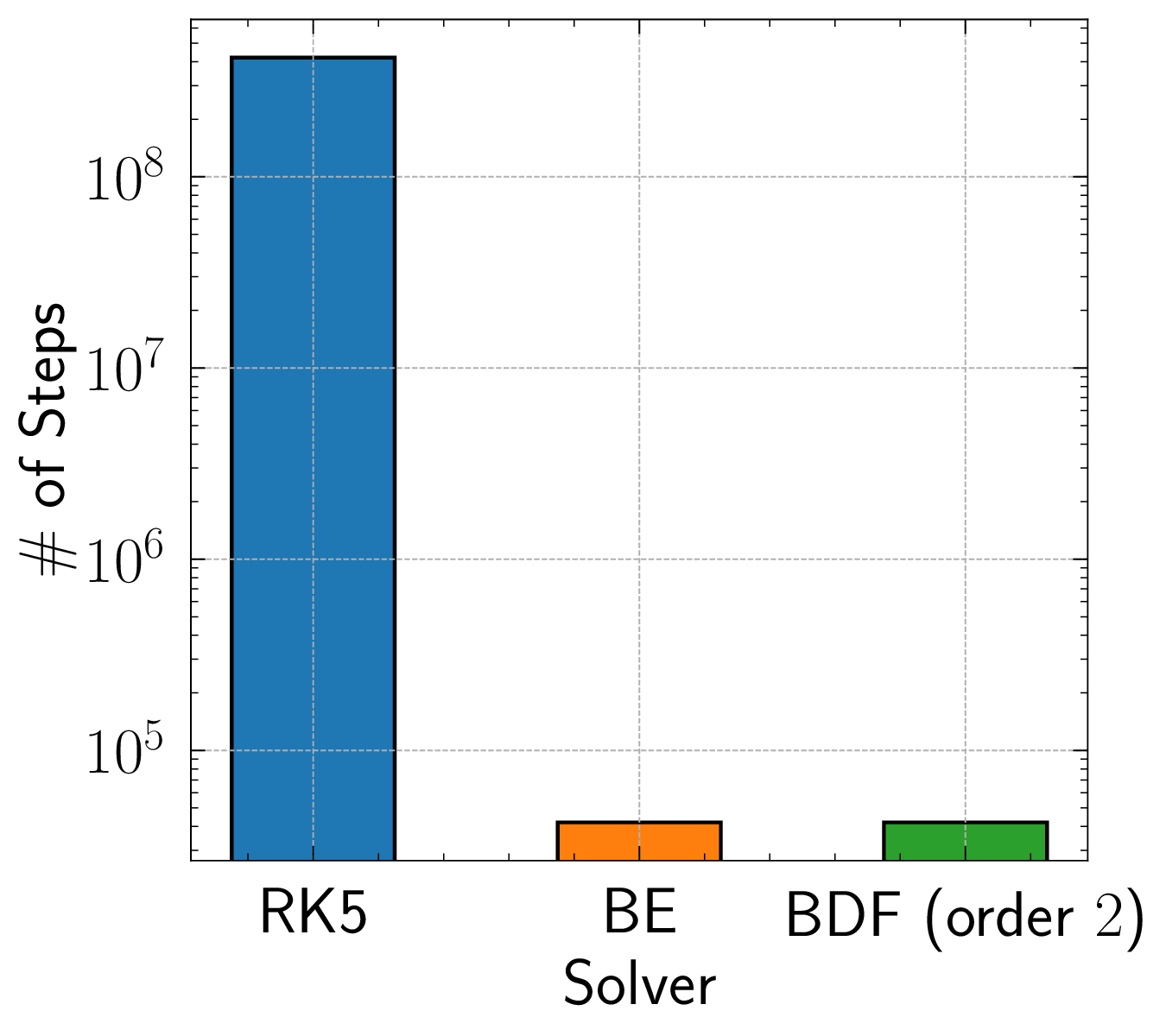}
        \caption{$\dot{\lambda} = 0.01/s$}
    \end{subfigure}\begin{subfigure}[H]{0.33\textwidth}
        \centering
        \includegraphics[width=\linewidth]{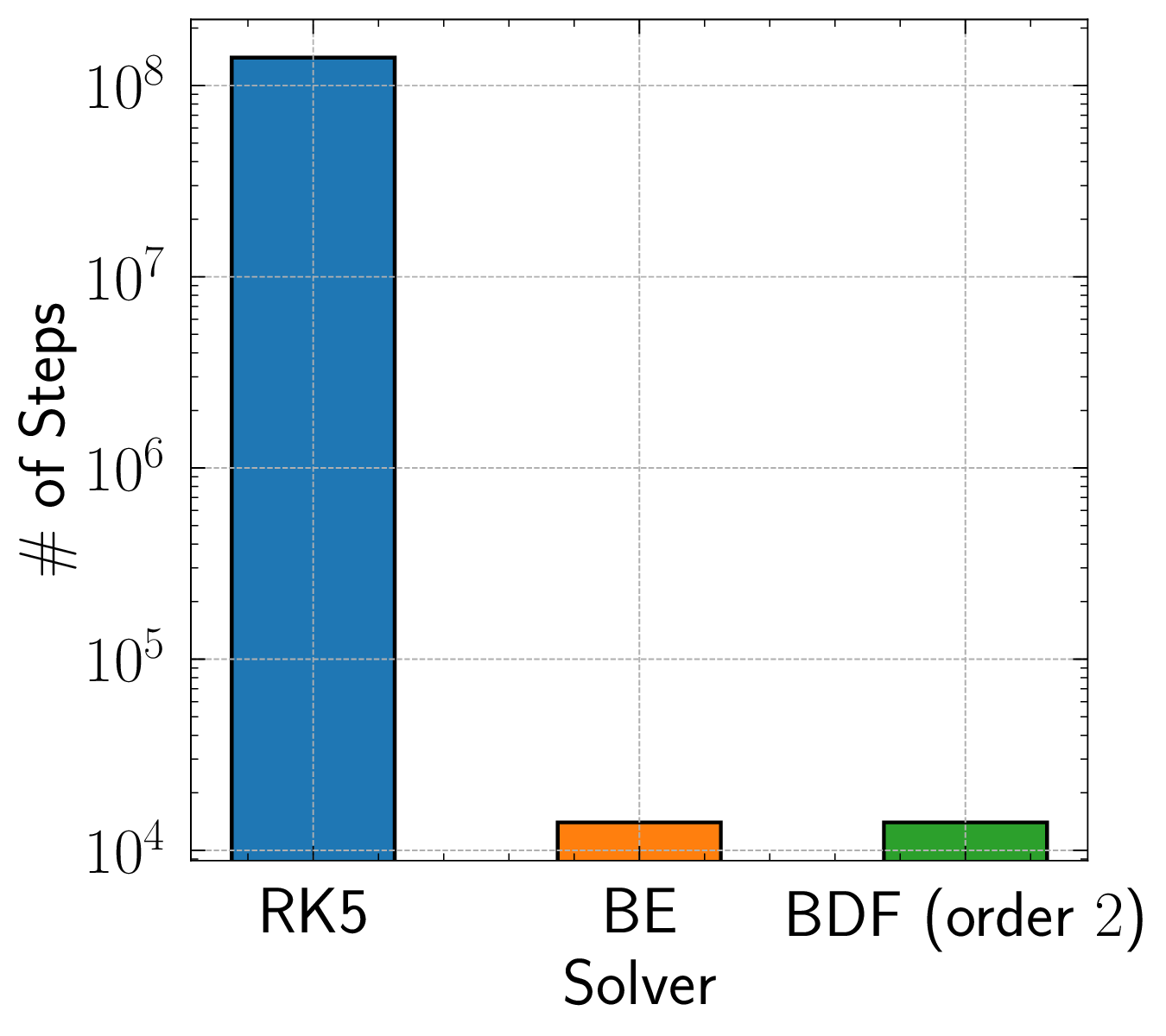}
        \caption{$\dot{\lambda} = 0.03/s$}
    \end{subfigure}\begin{subfigure}[H]{0.33\textwidth}
        \centering
        \includegraphics[width=\linewidth]{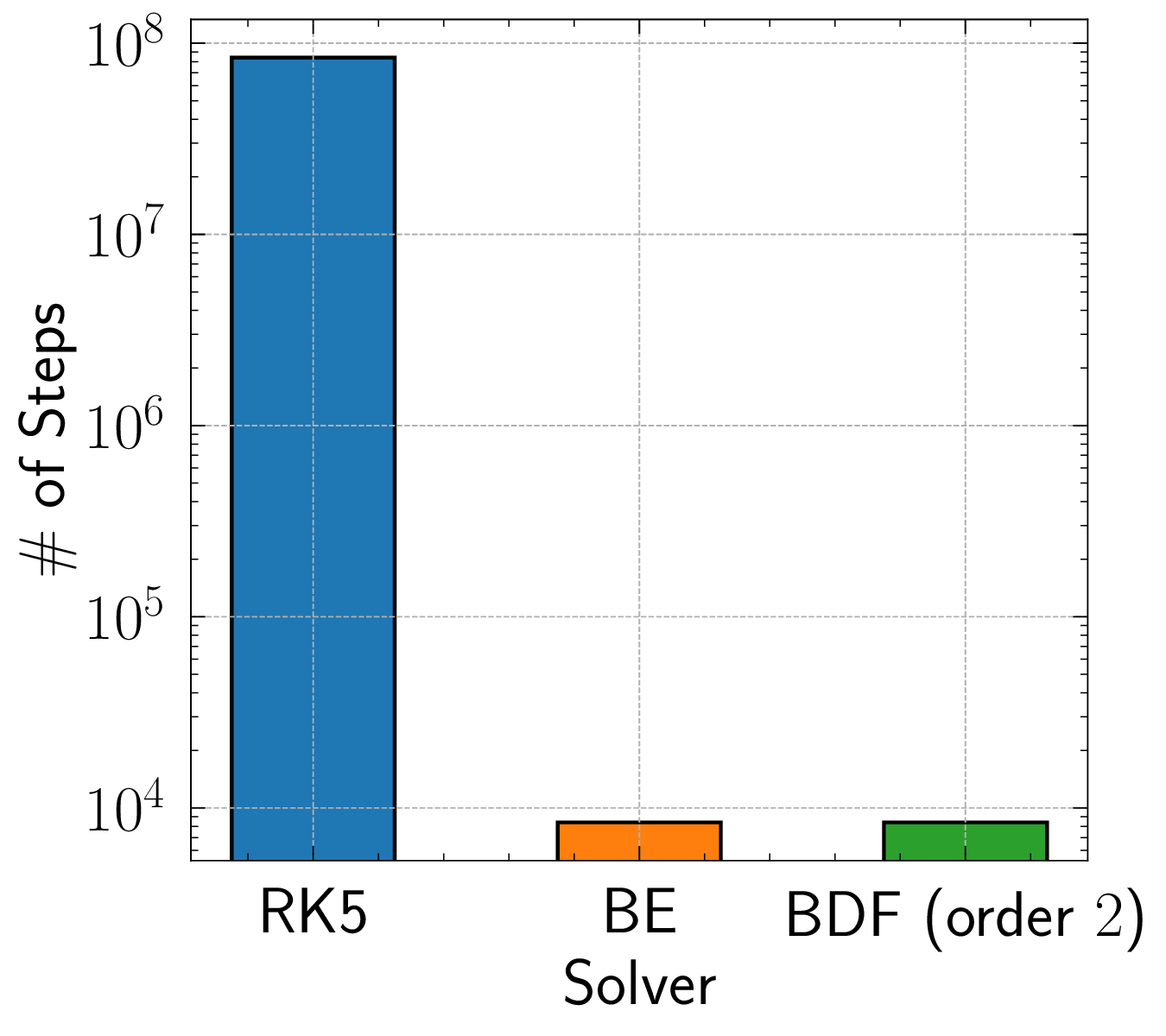}
        \caption{$\dot{\lambda} = 0.05/s$}
    \end{subfigure}
    \caption{The number of time steps taken by each solver to reach converged solutions corresponding to the loadings in Fig.~\ref{fig:applied_Stretch_ElecField_triangle} for VHB 4910. Note that RK5 takes significantly longer number of time-steps than Backward Euler (BE). This is also reflected in the total-time-to-solution (TTS) where BE significantly outperforms RK5 even with the additional cost of solving a nonlinear algebraic equation at each time-step. BDF (order 2) takes roughly the same number of time-steps to converge to the final solution.}\label{fig:num_steps_solvers}
\end{figure}
\begin{figure}[h!]
    \centering
    \begin{subfigure}[H]{0.33\textwidth}
        \centering
        \includegraphics[width=\linewidth]{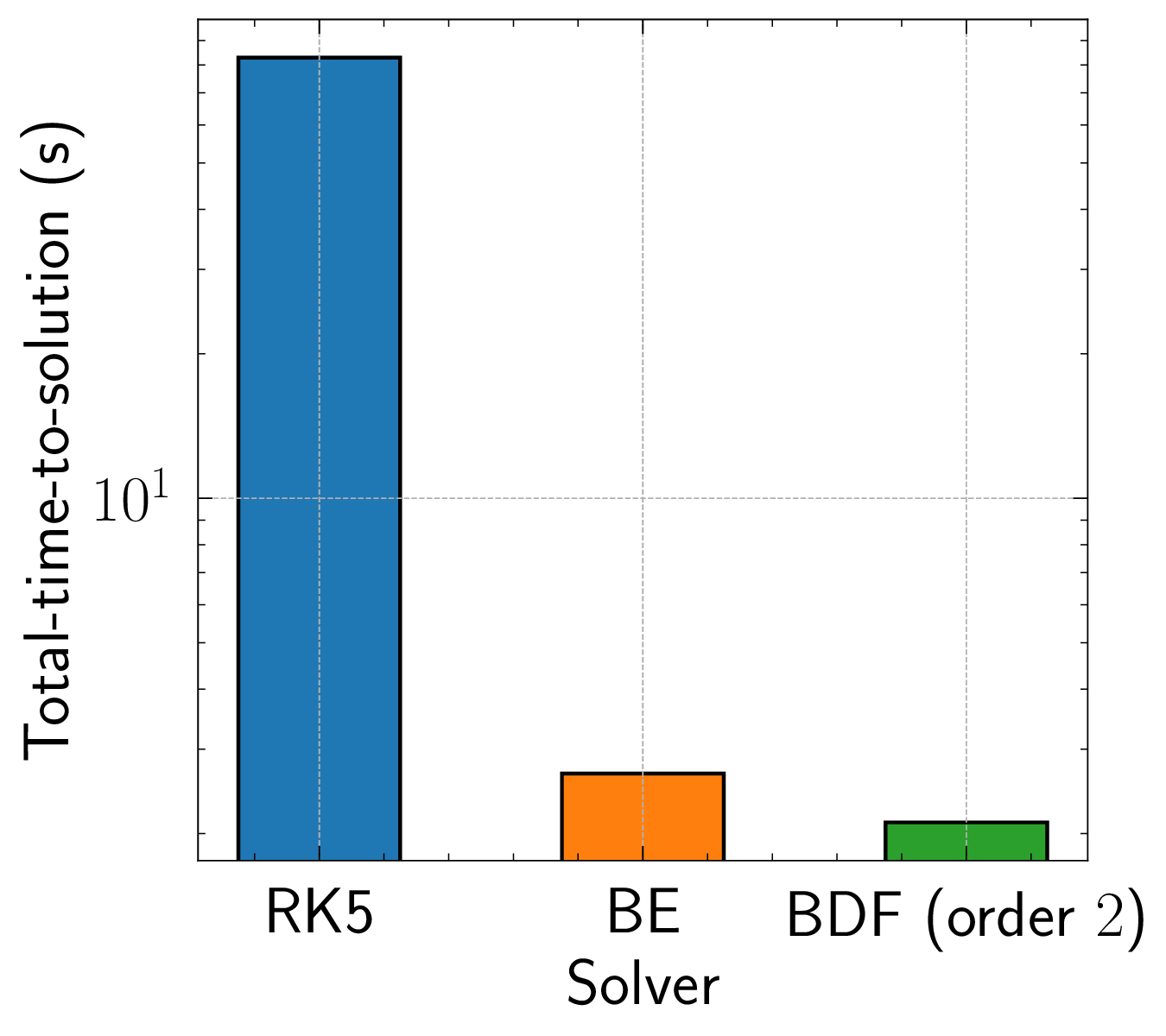}
        \caption{$\dot{\lambda} = 0.01/s$}
    \end{subfigure}\begin{subfigure}[H]{0.33\textwidth}
        \centering
        \includegraphics[width=\linewidth]{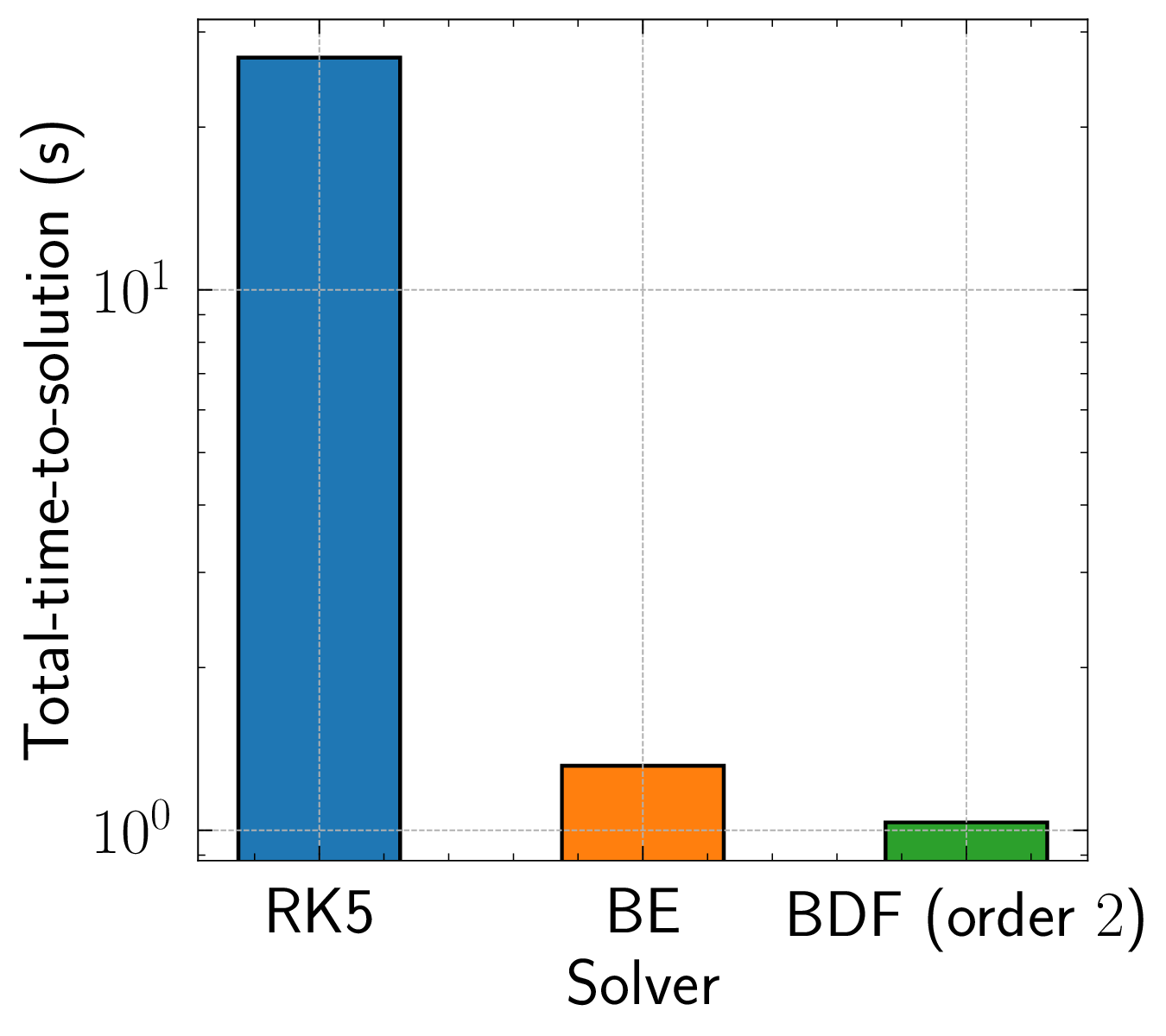}
        \caption{$\dot{\lambda} = 0.03/s$}
    \end{subfigure}\begin{subfigure}[H]{0.33\textwidth}
        \centering
        \includegraphics[width=\linewidth]{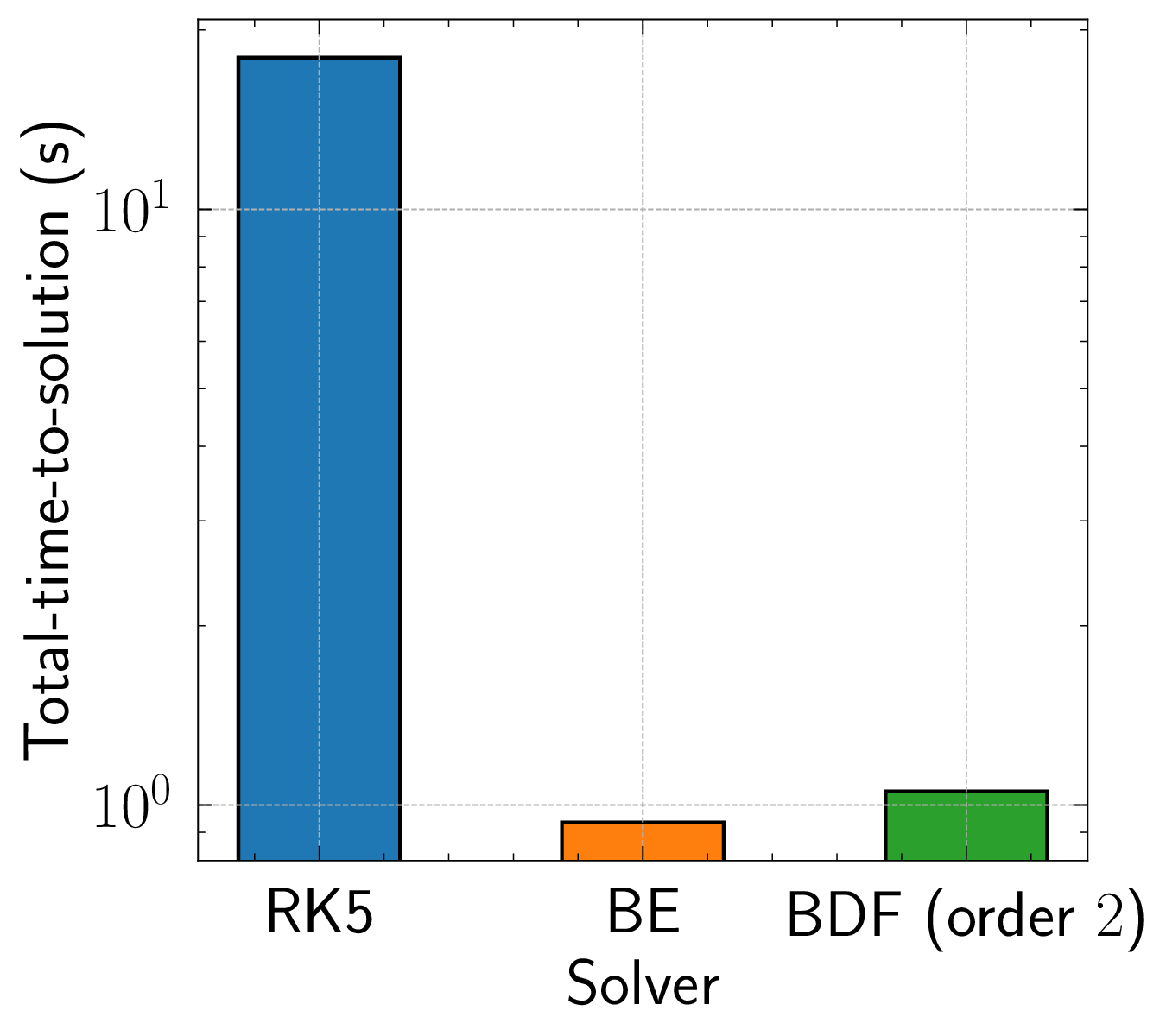}
        \caption{$\dot{\lambda} = 0.05/s$}
    \end{subfigure}
    \caption{The total-time-to-solution (TTS) in seconds for each time-integration algorithm corresponding to the loadings \eqref{eq:appliedDeformations_elecF}. RK5 takes significantly longer (an order of magnitude more) to converge to the final solution than both Backward Euler and BDF (order 2). Both Backward Euler and BDF reach the final converged solution in roughly the same time with the slight differences being possibly to better code-generation via \texttt{Numba}}\label{fig:tts_solvers}
\end{figure}
\begin{figure}[ht!]
    \centering
    \begin{subfigure}[H]{0.33\linewidth}
        \centering
        \includegraphics[width=\linewidth]{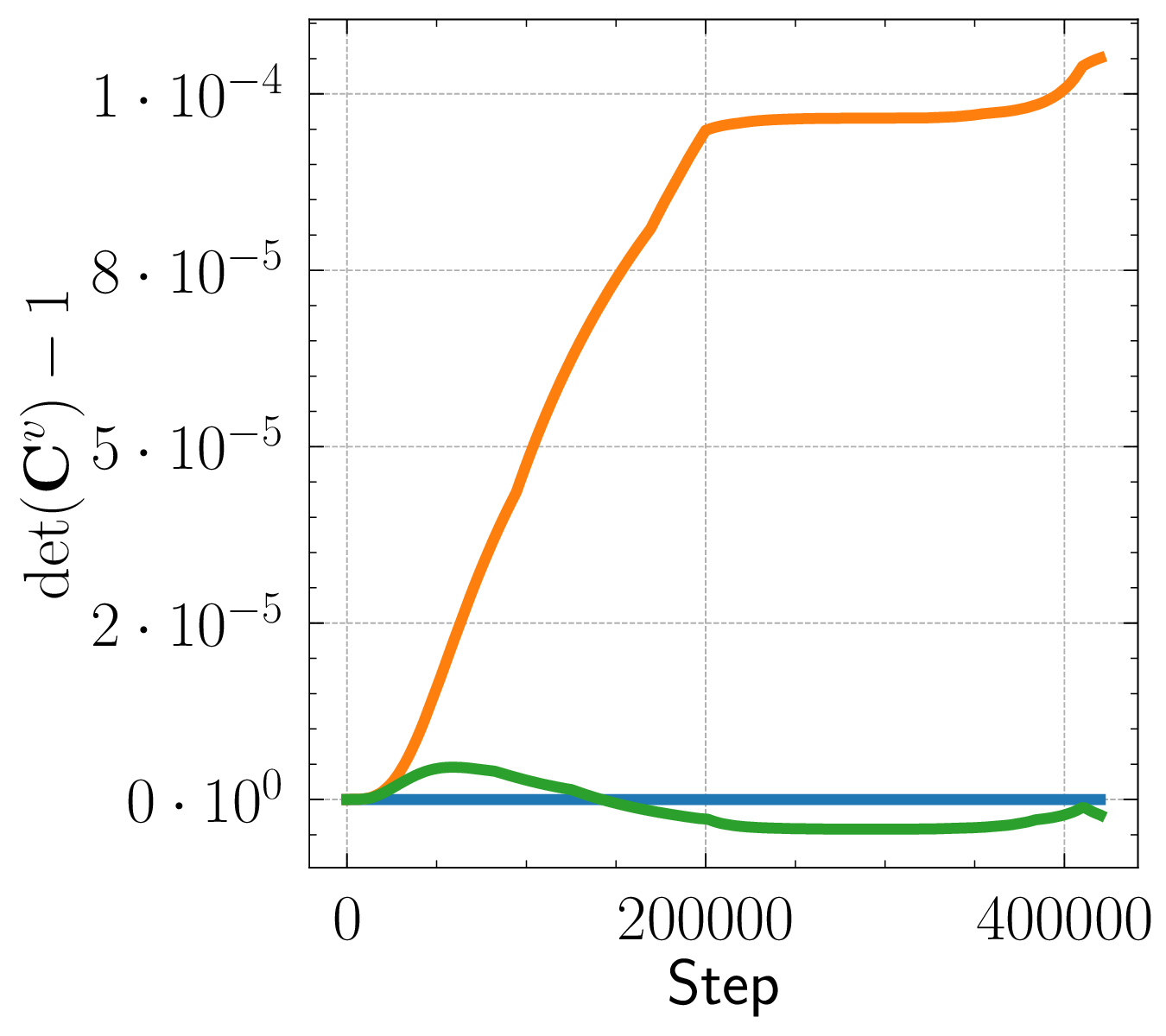}
        \caption{$\dot{\lambda} = 0.01/s$}
    \end{subfigure}\hfill\begin{subfigure}[H]{0.33\textwidth}
        \centering
        \includegraphics[width=\linewidth]{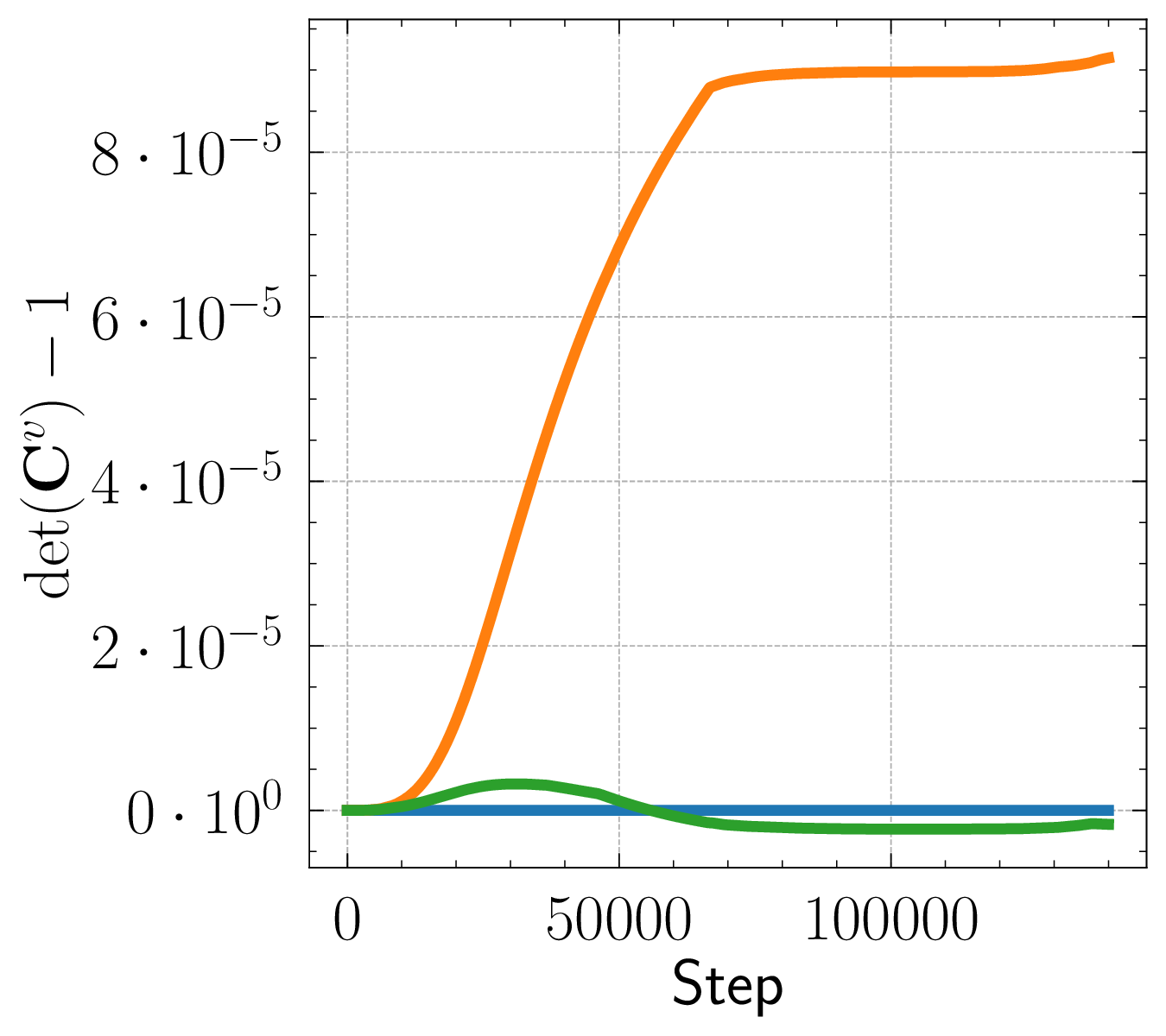}
        \caption{$\dot{\lambda} = 0.03/s$}
    \end{subfigure}\hfill\begin{subfigure}[H]{0.33\textwidth}
        \centering
        \includegraphics[width=\linewidth]{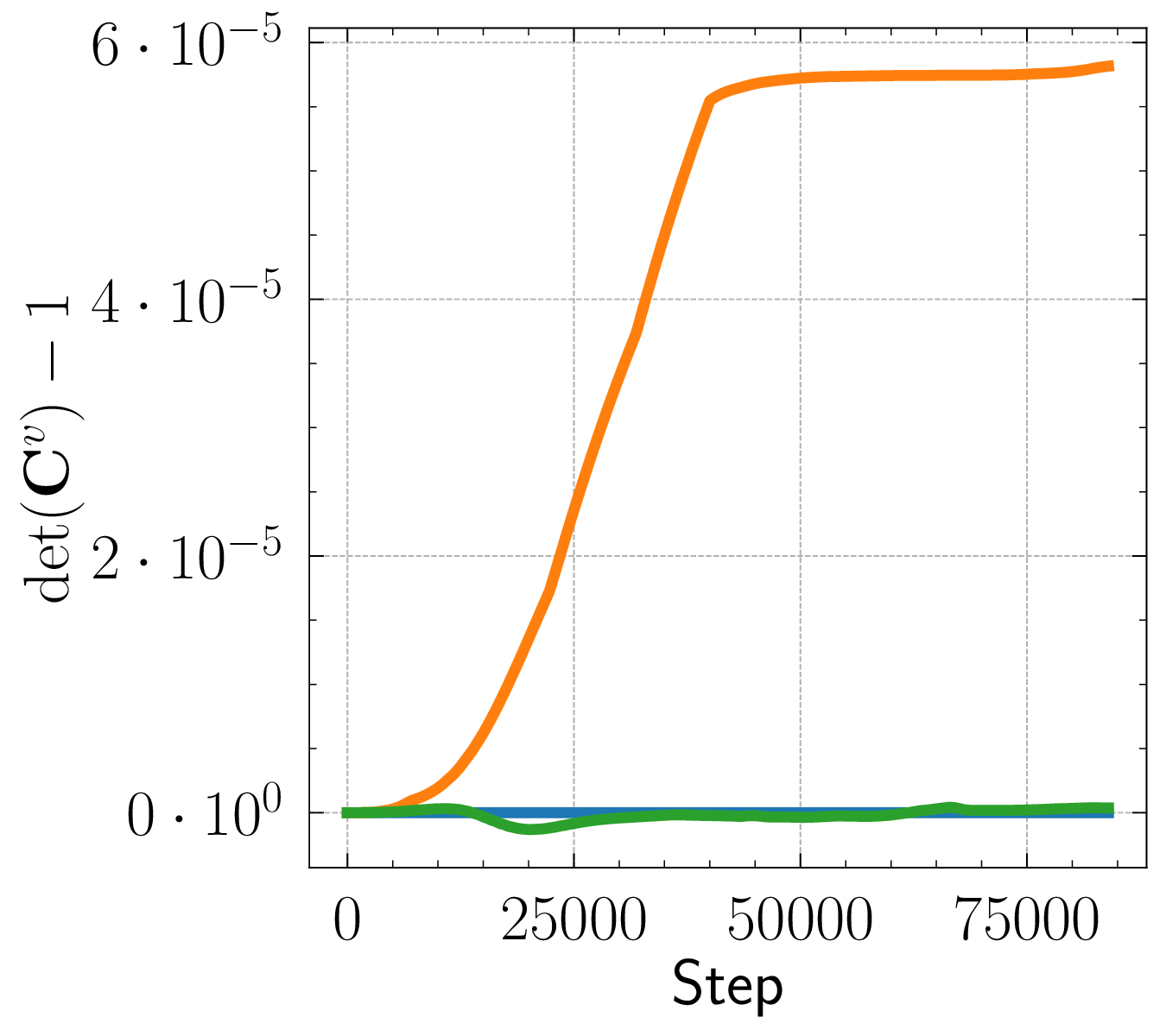}
        \caption{$\dot{\lambda} = 0.05/s$}
    \end{subfigure}\\
    \begin{subfigure}[H]{0.7\textwidth}
        \centering
        \includegraphics[width=\linewidth]{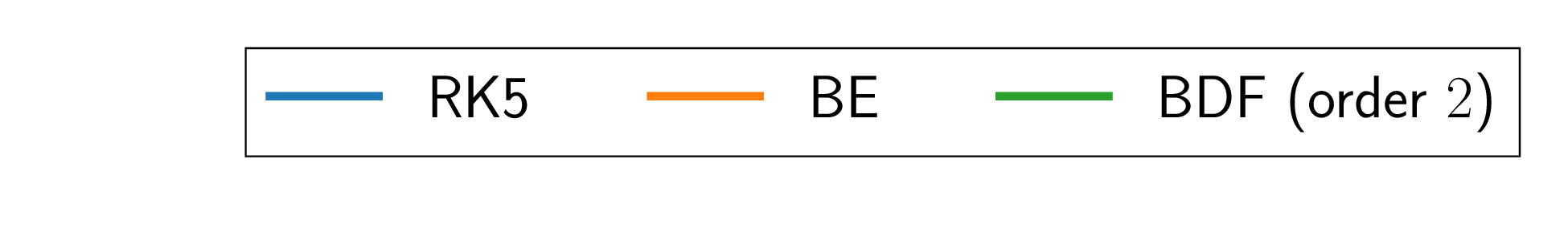}
        \caption*{}\vspace*{-4ex}
    \end{subfigure}
    \caption{Violation of the constraint ($\det(\mathbf{C}^v) = 1$) over the time-steps for the given loading \eqref{eq:appliedDeformations_elecF} for VHB4910. Note that RK5 being $5^{th}$-order accurate approximates the constraint the best, followed by BDF and BE.}\label{fig:det_constraint}
\end{figure}
\begin{remark}[Comparison between the different time-integration methods]
    For VHB 4910, all three solvers give nearly identical uniaxial stress-stretch response for both the electro-mechanical and mechanical loadings considered. We verify that the response is fully converged by decreasing the time-increment ($\Delta t\rightarrow 0$) and comparing the solutions. 
\end{remark}
\begin{remark}[Stable time-increment ($\Delta t$) for the different time-integration methods]
    The stable time-increment for Backward Euler (an implicit method) lies around $0.01$ seconds for the range of loadings considered. Similarly, for the $5$\textsuperscript{th} order RK solver it around $\ 10^{-6}$ seconds, around 4 orders of magnitude smaller than Backward Euler. For BDF, the stable time-increment lies in the range ($10^{-4}$, $10^{-3}$). 
\end{remark}
\begin{remark}[Total time-to-solution (TTS) and cost]
    The stable time increment has a direct influence on the time it takes to solve the initial value problems in \eqref{eq:ODES_Cv_Ev}. Furthermore, it also dictates the cost associated with the FE solver presented in Section~\ref{Sec:FEniCS-X}. The smaller the stable-time increment the larger number of steps required to reach the final solution and in turn the larger the cost of time-integration. We will remark on the cost associated with the FE solver in Section~\ref{Sec:FEniCS-X}. For the nonlinear ODEs in \eqref{eq:ODES_Cv_Ev}, we also present the total number of time-steps that each of the solvers take in Fig.~\ref{fig:num_steps_solvers} and the TTS in Fig.~\ref{fig:tts_solvers}. 
\end{remark}
\begin{remark}[Preserving the constraint $\det(\mathbf{C}^v) = 1$]
    None of the solvers presented in Section~\ref{sec:BE_ODE}-\ref{sec:BDF} preserve the constraint $\det(\mathbf{C}^v) = 1$ by design. While the high accuracy of RK5 is likely to preserve the constraint more closely, it too suffers from the same problem, albeit at a smaller resolution compared to the other solvers (see, e.g., the Appendix in \cite{shrimali2023mechanics}). We also report the violation of the constraint for VHB4910 and the loadings considered above in Fig.~\ref{fig:det_constraint}.
\end{remark}
\begin{remark}[RK5 as a suitable choice to integrate \eqref{eq:ODES_Cv_Ev}]
     Based on the results presented in Figs~(\ref{fig:stress_stretch_method_compare})-(\ref{fig:det_constraint}), we note that RK5 is not the best choice for integrating evolution equations arising in the coupled-electro-mechanical response of elastomers such as VHB 4910. Due to its high cost and TTS, it is often favorable to use lower-order implicit solvers such as BE. These results suffice to contradict what has previously been reported in the literature \cite{ghosh2021two}.
\end{remark}
% \begin{remark}[Modeling electric dissipation]\label{remark:modeling_electric_dissipation}
%     As illustrated in Fig.~\ref{fig:response_Ev_vs_noEv}, for homogeneous monotonic loadings the stress-stretch response is practically independent of the evolution of the internal variable $\bfE^v$ that models the electric dissipation. This is because the relaxation time for electrical dissipation is much smaller than the mechanical relaxation. However, this may not be true for arbitrary loadings in time, such as in  \cite{qiang2012experimental}, where indeed the electric dissipation may need to be explicitly modeled in order to determine the electromechanical response of the dielectric elastomer of interest. However, since we only consider monotonic loadings in this paper, we ignore the evolution of $\bfE^v$ in the numerical results presented in Section.~\ref{Sec:FEniCS-X}.
% \end{remark}

\section{3D Implementation: FEniCS-X}\label{Sec:FEniCS-X}
\subsection{FE Formulation}
For arbitrary boundary conditions, loadings and geometries we need to solve the governing PDEs \eqref{BVP-F-hybrid}-\eqref{BVP-E-hybrid} together with \eqref{Evol-Eq-Cv3} and \eqref{Evol-Eq-Ev3} for the internal variables. In this section we present a robust numerical framework that makes use of an implicit FE scheme to discretize the deformation field ($\bfy$), the hydrostatic pressure ($p$), the electric potential ($\varphi$) together with a Backward-Euler discretization in-time of the evolution equations \eqref{Evol-Eq-Cv3}.
% and \eqref{Evol-Eq-Ev3}.

% \subsection{3-D FE Implementation} \label{3-D_Hybrid_Problem}
To this end, we spell out the corresponding Galerkin form of \eqref{BVP-F-hybrid-weak} together with an implicit discretization for the evolution equations \eqref{Evol-Eq-Cv3} and its numerical implementation within the open-source FE framework \texttt{FEniCSx}~\cite{baratta2024dolfinx}. We discretize the fields $\bfy^h_{n+1}(\bfX)$, $p^h_{n+1}(\bfX)$ and $\phi^h_{n+1}(\bfX)$ at time $t_{n+1}$ using piecewise conforming finite elements of degree 2, 1 and 2 respectively, i.e. 
\begin{align}
    \begin{aligned}
        \bfy^{h}_{n+1}|_{\mathcal{E}^e} &\subset \boldsymbol{\mathcal{P}}^2 (\mathcal{E}^e)\\
    p^{h}_{n+1}|_{\mathcal{E}^e} &\subset {\mathcal{P}}^1 (\mathcal{E}^e)\\
    \phi^{h}_{n+1}|_{\mathcal{E}^e} &\subset {\mathcal{P}}^2 (\mathcal{E}^e),
    \end{aligned}\label{eq:FE_spaces}
\end{align}
where $\bfy_{n+1}(\bfX)\equiv(\bfy^h(\bfX, t_{n+1})$, $\phi_{n+1}(\bfX)\equiv(\phi^h(\bfX, t_{n+1})$, $p_{n+1}(\bfX)\equiv(p^h(\bfX, t_{n+1})$ and $(\cdot)|_{\mathcal{E}^e}$ represents the restriction on element $\mathcal{E}^e$. Furthermore, $\mathcal{P}^k(\mathcal{E}^e)$ and $\boldsymbol{\mathcal{P}}^k(\mathcal{E}^e)$ correspond to $k$\textsuperscript{th} degree scalar and vector lagrange elements on $\mathcal{E}^e$ respectively.
Note that this combination of FE spaces \eqref{eq:FE_spaces} corresponds to the lowest order analogue of the popular Taylor-Hood elements (see, e.g., Chapter VI in \cite{brezzi2012mixed}) in incompressible elasticity. Having spelled out the weak form and the corresponding FE spaces for the fields, we now turn to the Galerkin form at time $t_{n+1}$. Let $\texttt{N}_e$ denote the total number of elements in the FE mesh of $\Omega$, $\texttt{N}_t$ denote the set of elements whose boundaries lie on $\partial \Omega_0^{\mathcal{N}}$ and $\texttt{N}_{t_{\varphi}}$ denote the set of elements whose boundaries lie on $\partial \Omega_0^{\mathcal{N}_{\varphi}}$, then the corresponding Galerkin form is given by
\begin{align}
    \hspace*{-3ex}
    \left\{\begin{array}{ll}
    \displaystyle
     \underset{e=1}{\overset{\texttt{N}_e}{\mathbb{A}}}
     \int_{\mathcal{E}^e}
     \left[\bfS_{n+1}\left(\nabla\bfy^h_{n+1}, -\nabla\varphi^h_{n+1},\bfC^v_{n+1},\bfE^v_{n+1}\right) 
     + 
     p_{n+1}J_{n+1} \nabla \bfy_{n+1}^{-T}\right] \cdot \nabla \mathbf{v} \ \rm d \bfX
     =\\[10pt]
     \qquad \qquad \displaystyle
     \underset{e=1}{\overset{\texttt{N}_t}{\mathbb{A}}}
     \int_{\mathcal{E}^e}
     \mathbf{f}_{n+1}(\bfX)\cdot \mathbf{v} \ \rm d \bfX
     + 
     \underset{e=1}{\overset{\texttt{N}_t}{\mathbb{A}}}
     \int_{\partial \mathcal{E}^e}\overline{\textbf{t}}_{n+1}(\bfX)\cdot \mathbf{v} \ \rm d \bfX, \quad \forall \mathbf{v} \in \mathcal{Y}_0\\[10pt]
    \displaystyle
    \underset{e=1}{\overset{\texttt{N}_e}{\mathbb{A}}}
    \int_{\mathcal{E}^e}
    \left(\det\nabla\bfy_{n+1}(\bfX) 
    -\dfrac{1 + p_{n+1}/\kappa+\sqrt{4\sum_{r=1}^2\mu_r/\kappa + (1+p_{n+1}/\kappa)^2}}{2}\right)q \ \rm d \bfX = 0,\quad \forall \textrm{$q$}\in \mathcal{P}_0\\[10pt]
     \displaystyle
     \underset{e=1}{\overset{\texttt{N}_t}{\mathbb{A}}}
     \int_{\mathcal{E}^e}{\rm Div}\left[\bfD_{n+1}\left(\nabla\bfy_{n+1},-\nabla\varphi_{n+1},\bfC_{n+1}^v,\bfE_{n+1}^v\right)\right]\cdot \phi \ \rm d \bfX = \\[10pt]
     {\qquad \qquad \displaystyle\underset{e=1}{\overset{\texttt{N}_e}{\mathbb{A}}}\int_{\mathcal{E}^e}Q_{n+1}(\bfX)\phi \ \rm d \bfX 
     + 
     \displaystyle
     \underset{e=1}{\overset{\texttt{N}_{t_{\varphi}}}{\mathbb{A}}}
     \int_{\partial \mathcal{E}^e}\left(\sigma_{n+1}(\bfX)
     -
     \overline{\bfD}_{n+1}(\bfX)\cdot\bfN\right) \ \rm d \bfX,\quad \forall \textrm{$\phi$}\in \mathcal{V}_0,}
     % _{\displaystyle\Pi^h\left(\bfy_{n+1}, p_{n+1}, \varphi_{n+1};\bfC^v_{n+1}, \bfE^v_{n+1}\right)=0}
     % \underbrace_{{\Pi^h}}
     \end{array}
     \right. \label{BVP-F-hybrid-galerkin}
\end{align}
where the internal variables $\bfC^{v^h}_{n+1}$ and $\bfE^{v^h}_{n+1}$ are given by the solution of
\begin{align}
    \begin{aligned}
        \bfC^{v^h}_{n+1} = \mathbf{G}(t_{n+1}, \bfC^{v^h}_{p}; \bfy^h_{n+1}),\qquad \text{where}\qquad p\in \{n,n+1\},
    \end{aligned}\label{eq:CV_evol_eq_quadPoints}
\end{align}
and
\begin{align}
    \begin{aligned}
        \bfE^{v^h}_{n+1} = \mathbf{H}(t_{n+1}, \bfE^{v^h}_{p}; \bfy^h_{n+1}, \varphi^h_{n+1}),\qquad \text{where}\qquad p\in \{n,n+1\},
    \end{aligned}\label{eq:EV_evol_eq_quadPoints}
\end{align}
at each quadrature point in $\mathcal{E}$ respectively. 
\begin{remark}[Backward-Euler as the time-integration algorithm]
    It is clear from the results presented in Section~\ref{subsec:1D_problem_Numerics}, specifically those in Figure~\ref{fig:tts_solvers} that Backward-Euler is a suitable choice to integrate the evolution equation and therefore we choose $p=n+1$ in \eqref{eq:CV_evol_eq_quadPoints}.
\end{remark}
% Note that the nonlinear form \ref{BVP-F-hybrid-galerkin} contains the internal variable $\bfE^v$ that describes the electric dissipation. While this is important for certain classes of loadings (see, e.g. spectroscopy experiments in \cite{qiang2012experimental}), for monotonic loadings, $\bfE^v=\bfE$ for all times ($t$). Therefore, we do not consider the evolution of $\bfE^v$ in our FE simulations. 
Finally, and the system of nonlinear equations can be written as 
\begin{equation}
    \Pi^h\left(\bfy_{n+1}, p_{n+1}, \varphi_{n+1};\bfC^v_{n+1}\right)=0.
\end{equation}
We can then compute the first order taylor-expansion about an arbitrary point $\bfw = (\bfy_n, p_n, \varphi_n)$ to get
\begin{equation}
    \Pi^h\left(
        \bfw_n;\bfC^v_{n+1}
    \right) + \dfrac{\mathrm{D}{\Pi}^h}{\mathrm{D}\bfw}\delta \bfw = 0 \implies \delta\bfw = \left[ 
        \dfrac{\Pi^h}{D\bfw}
    \right]^{-1} \cdot \Pi^h\left(
        {\bfw}_n;\bfC^v_{n+1}
    \right)
\end{equation}
which can be solver in conjunction with \eqref{eq:CV_evol_eq_quadPoints} and \eqref{eq:EV_evol_eq_quadPoints} for $\delta\bfw = \bfw_{n+1}-\bfw_n$ and in turn $\bfw_{n+1}$. The corresponding algorithm is outlined in Algorithm~\ref{alg:time_stepping}.
% \begin{remark}[Solving the evolution equation for the internal variable $\bfC^v_{n+1}$]
%     The system of nonlinear equations \eqref{eq:CV_evol_eq_quadPoints} stems from an implicit discretization of evolution equation $\dot{\bfC}^v = \mathbf{G}(t, \bfC^v;\mathbf{C})$, e.g. in a BE time discretization in Section~\ref{sec:BE_ODE}. This needs to be solved at each quadrature point in the FE mesh in combination with a global nonlinear solver for the fields. We elaborate more on this in Algorithm~\ref{alg:time_stepping}
% \end{remark}
% \begin{remark}[Solving the nonlinear form \eqref{BVP-F-hybrid-galerkin}]
%     The nonlinear form \eqref{BVP-F-hybrid-galerkin} requires a nonlinear solver (e.g. Newton's method), which requires computation of the jacobian, which comprises of the following computations: $\partial\bfS/\partial\bfF$, $\partial\bfS/\partial p$ and $\partial\bfS/\partial\\varphi$.
% \end{remark}

\begin{figure}[H]
    \centering
    \includegraphics[width=\linewidth]{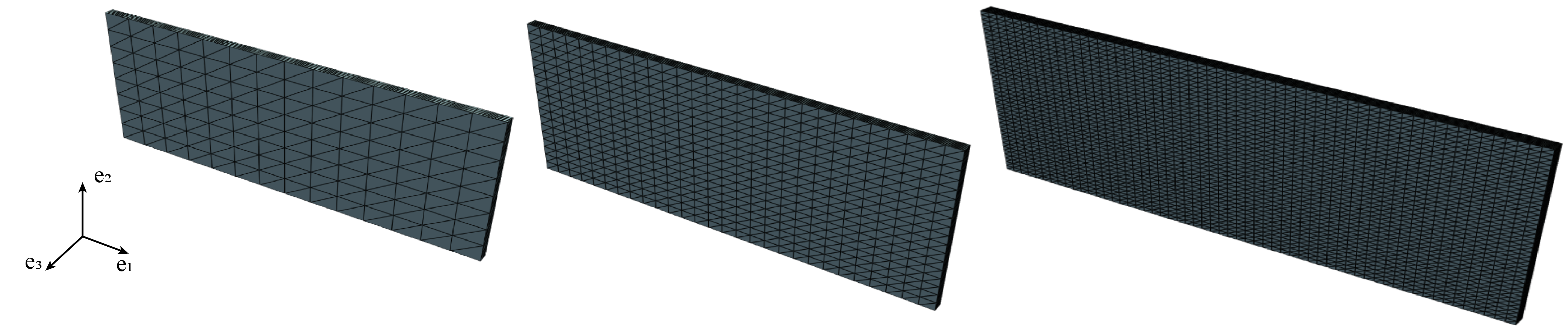}
    \caption{Finite-element meshes for the FEniCSx implementation used in the study. We consider a sequence of 3 meshes comprised of quadratic (10-noded) tetrahedra with uniform refinement, ranging from $\sim$1000 elements (left) to about $\sim$80,000 elements (right).}\label{fig:FE_meshes}
\end{figure}
Specifically, discretize the displacement, pressure and electric potential using conforming FE element spaces. 

\begin{algorithm}[H]
    \caption{Algorithm for the nonlinear solver for the hybrid problem \eqref{BVP-F-hybrid-galerkin}-\eqref{eq:CV_evol_eq_quadPoints}}\label{alg:time_stepping}
    \begin{algorithmic}[1]
    \STATE {\textbf{Given:} Dirichlet boundary conditions for the deformation field $\overline{\mathbf{y}}(\mathbf{X}, t)$ and voltage ($\overline{\phi}(\bfX, t)$)}
    \FOR{$t \in \{0, T_{\text{final}}\}$}
        \STATE Get boundary conditions at time $t_{n+1}$, the DOFs at time $t_n$ i.e., $(\mathbf{u}_n, p_n, \phi_n)$, the internal variable $\mathbf{C}^v_n(\mathbf{X})$ and initialize the global DOF-vector $(\mathbf{u}, p, \phi)$
        \FOR{$k \in \{1, \texttt{max\_newton\_iter}\}$}
            \STATE Initialize global (sparse) stiffness matrix $\boldsymbol{K}$ and force vector ($\boldsymbol{f}$)
            \FOR{element $e \in \{1, \ldots, \texttt{N}^e\}$}
                \STATE Initialize local stiffness matrices $\boldsymbol{k}^{(e)}_{\mathbf{u}\mathbf{u}}$, $\boldsymbol{k}^{(e)}_{\mathbf{u}p}$, $\boldsymbol{k}^{(e)}_{\mathbf{u}\phi}$ and force vectors ($\boldsymbol{f}^{(e)}$)
                \FOR{\texttt{qp} $\in \{1, \ldots, \texttt{num\_quad\_points}\}$}
                    \STATE {Compute ($\mathbf{C}(\mathbf{X}_{\texttt{qp}})$) from displacement ($\mathbf{u}(\mathbf{X}_{\texttt{qp}})$) at \texttt{qp}}
                    \STATE {Solve \eqref{eq:CV_evol_eq_quadPoints} for $\mathbf{C}^v_{n+1}(\mathbf{X}_{\texttt{qp}})$ using Newton's method, and assign}
                \ENDFOR
                \STATE $\boldsymbol{k}^{(e)} = \texttt{assemble}(\boldsymbol{k}^{(e)}_{\mathbf{u}\mathbf{u}}, \boldsymbol{k}^{(e)}_{\mathbf{u}p}, \boldsymbol{k}^{(e)}_{\mathbf{u}\phi})$
                \STATE $\boldsymbol{f}^{(e)} = \texttt{assemble}(\mathbf{S}_{n+1,k})$
            \ENDFOR
        \ENDFOR
    \ENDFOR
    \end{algorithmic}
\end{algorithm}

% \{Governing equations}
\subsection{Implementation and Results}
In this section we present the finite-element (FE) implementation of the governing equations \eqref{BVP-F-hybrid-weak} using the open-source FE library \texttt{dolfinx}~\cite{baratta2024dolfinx}. We implement the weak form using the unified form language \texttt{UFL}~\cite{alnaes2014unified}, and make use of quadrature elements for the internal variable ($\mathbf{C}^v$) available in \texttt{basix}~\cite{scroggs2022basix}. We finally make use of \texttt{PETSc} (using the python wrappers in \texttt{petsc4py})~\cite{balay2022petsc} to solve the resulting system of nonlinear equations at each time-step.
\begin{figure}[h!]
\centering
    \begin{subfigure}[H]{\textwidth}
    \centering
        \includegraphics[width=\linewidth]{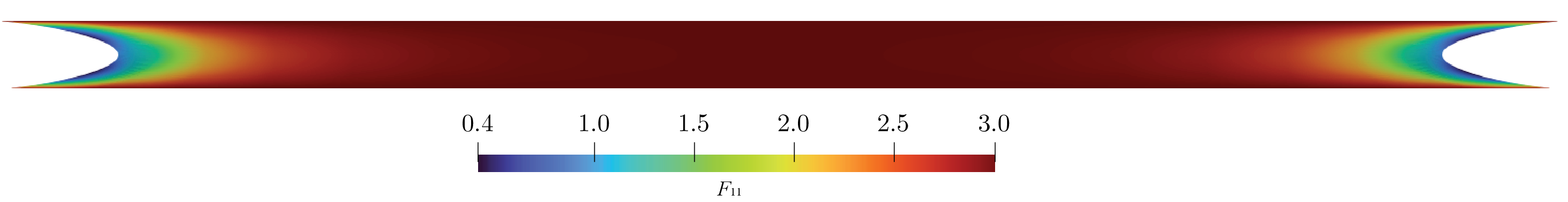}
        \caption{Stretch $\lambda_{1}(\mathbf{x})$ after pre-stretching the elastomeric specimen (see Fig. 3 in \cite{hossain2015comprehensive})}
    \end{subfigure}\\
    \begin{subfigure}[H]{\textwidth}
    \centering
        \includegraphics[width=\linewidth]{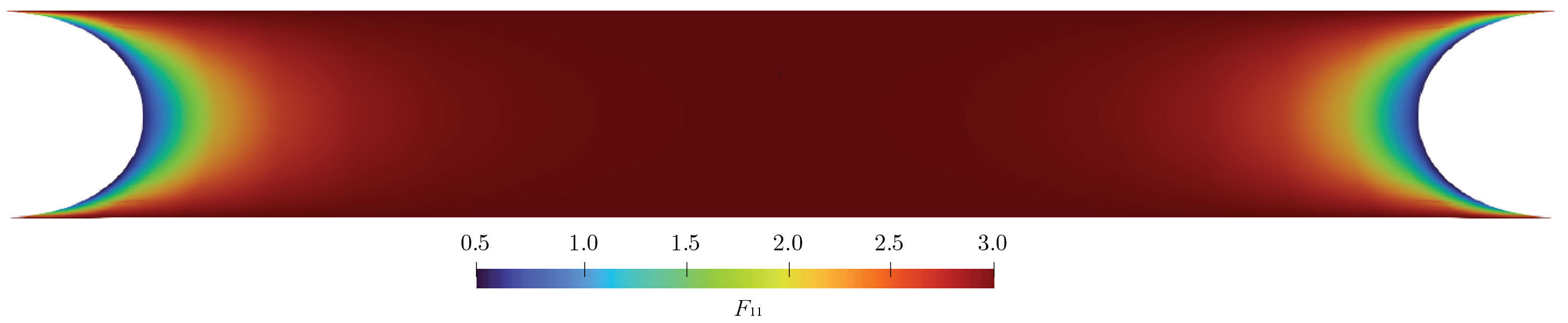}
        \caption{Stretch $\lambda_{1}(\mathbf{x})$ in the fully loaded configuration corresponding to $\lambda_2 = 2.25$.}
    \end{subfigure}
    \caption{Contours of the principal stretch $F_{11}(\mathbf{x})$ in (a) pre-stretched configuration, and (b) in full-loaded configuration. Note the non-homogeneous stretches at the boundaries}
\end{figure}

\begin{figure}[h!]
\centering
    \begin{subfigure}[H]{0.33\textwidth}
    \centering
        \includegraphics[width=\linewidth]{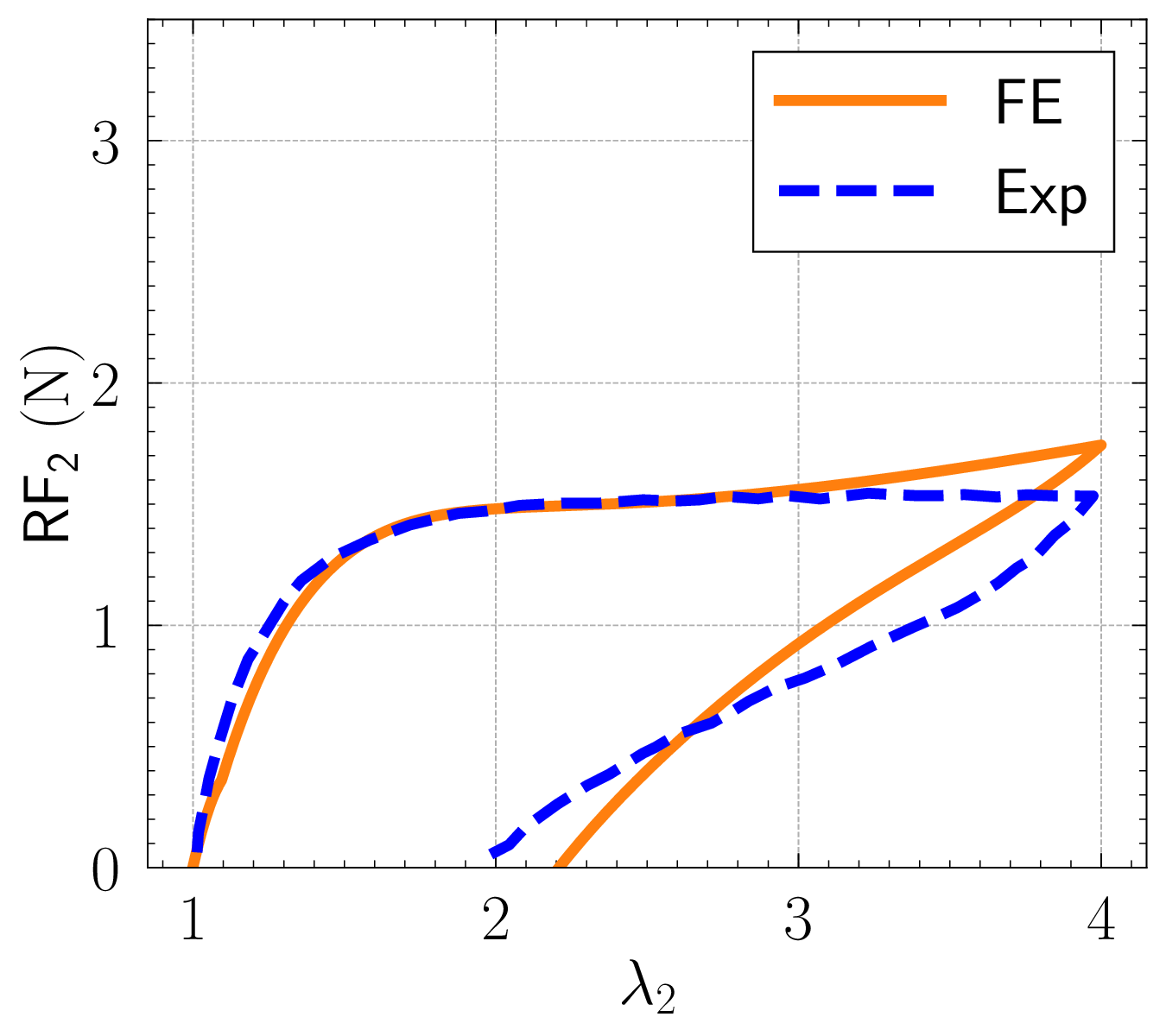}
        \caption{$\dot{\lambda}_2 = 0.01/s$}
    \end{subfigure}\begin{subfigure}[H]{0.33\textwidth}
    \centering
        \includegraphics[width=\linewidth]{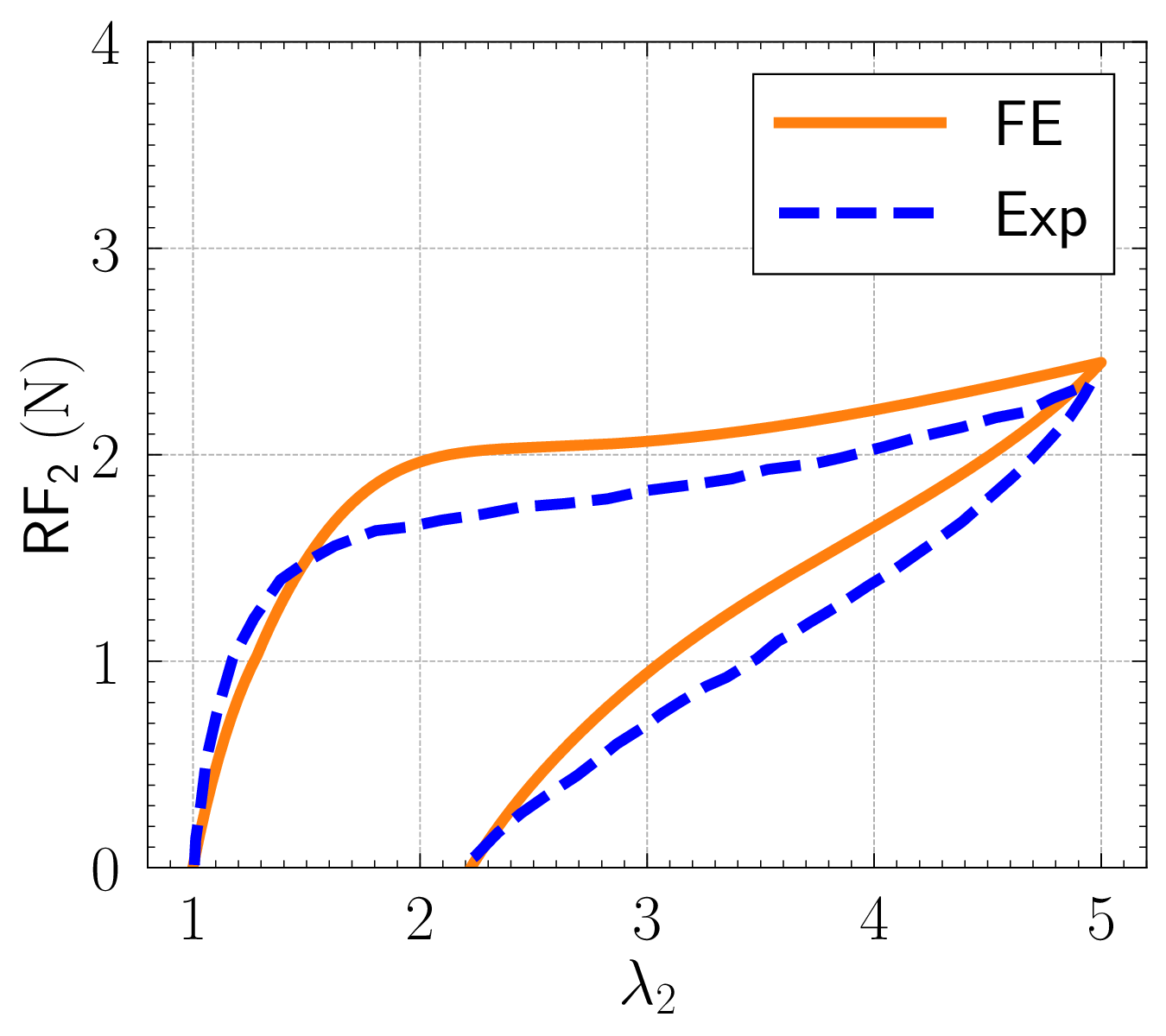}
        \caption{$\dot{\lambda}_2 = 0.01/s$, $\Delta\varphi = 4$~kV}
    \end{subfigure}\begin{subfigure}[H]{0.33\textwidth}
    \centering
        \includegraphics[width=\linewidth]{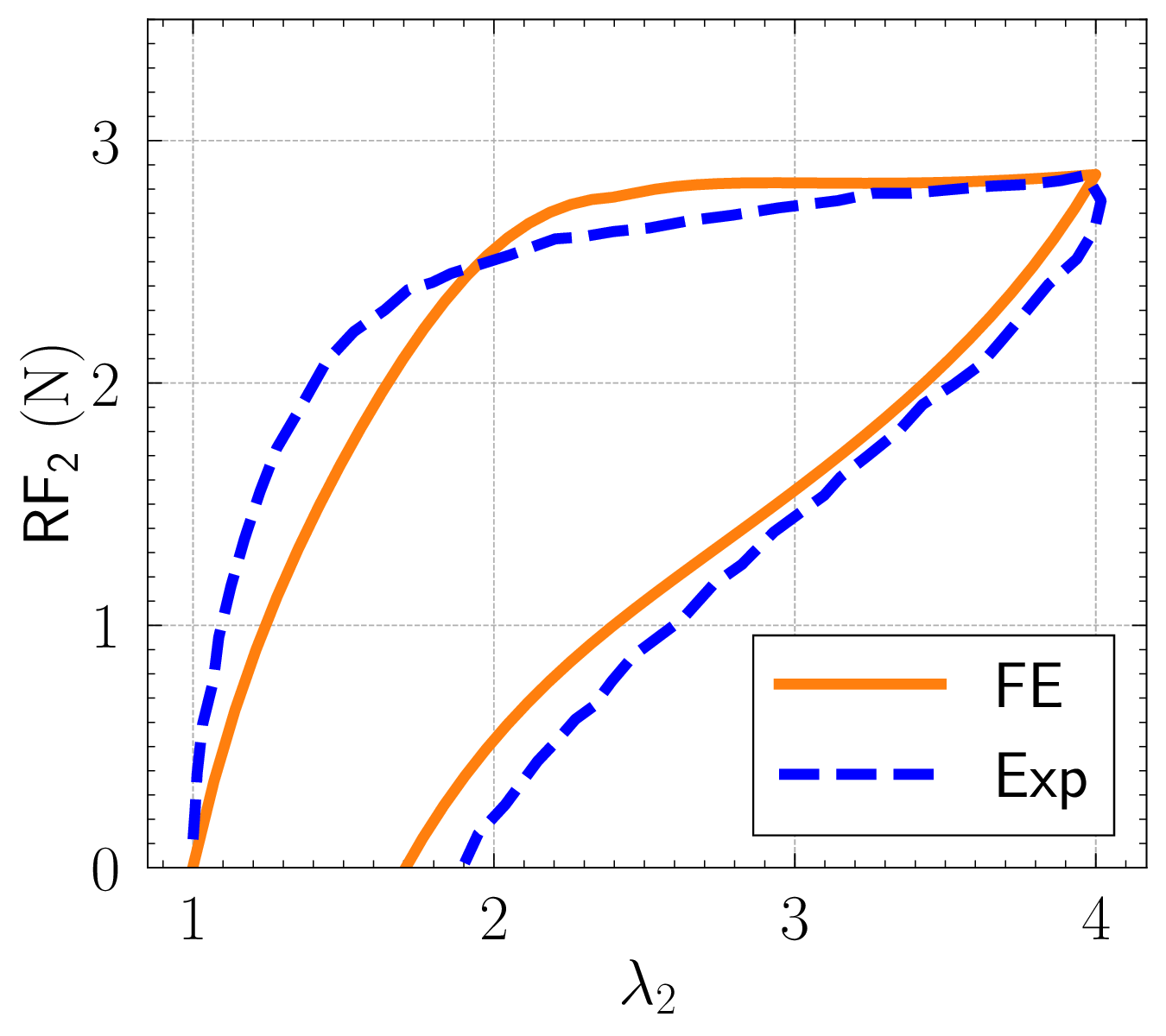}
        \caption{$\dot{\lambda}_2 = 0.05/s$}
    \end{subfigure}
    \caption{Comparisons with the experiments in \cite{hossain2015comprehensive}. Figures (a)-(c) show the reaction force in the $\bfe_2$ direction as a function of the applied stretch $\lambda_2$ for a VHB4910 specimen pre-stretched to 200\% deformation ($\lambda_1=2$) in the $\bfe_1$ direction and subjected to a constant potential difference ($\Delta\varphi$) in the $\bfe_3$ direction}
\end{figure}

\subsection{Defining constitutive behavior in FEniCS-X}
We provide snippets of the \texttt{python} code that define the constitutive behavior \eqref{Constitutive_Equations}
\begin{lstlisting}[language=Python,caption=Defining the ``total energy'' used for computing the weak form (residual) and its Jacobian (tangent modulus) in \texttt{dolfinx}]
from ufl import Identity, det, grad, tr, inner, ssqrt, ln

def Psi(u, p, phi, Cv, Ev, params):
    (mu1, mu2, alph1, alph2, m1, m2, a1, a2, K1, K2, bta1, bta2, eta0, etaInf, mk, nk, eps, veps, zeta, lam1) = params
    d = len(u)
    I = Identity(d)
    F = I + grad(u)
    C = F.T * F
    Cv_inv = inv(Cv)
    I1 = tr(C)
    J = det(F)
    E = -grad(phi)
    Ee = E - Ev
    I4 = inner(E, E)
    I5 = inner(E, inv(C) * E)
    Ie4 = inner(Ee, Ee)
    Ie5 = inner(Ee, inv(C) * Ee)
    Ie1 = inner(C, Cv_inv)
    Jv = sqrt(det(Cv))
    phat = lam1 + p
    Jstar = (phat + sqrt(phat**2 + 4 * lam1 * (mu1 + mu2 + m1 + m2))) / (2 * lam1)
    psi_Eq = ((3**(1 - alph1)) / (2 * alph1)) * mu1 * (I1**alph1 - 3**alph1) + ((3**(1 - alph2)) / (2 * alph2)) * mu2 * (I1**alph2 - 3**alph2) - (mu1 + mu2) * ln(Jstar) + 0.5 * lam1 * (Jstar - 1)**2
    psi_Neq = ((3**(1 - a1)) / (2 * a1)) * m1 * (Ie1**a1 - 3**a1) + ((3**(1 - a2)) / (2 * a2)) * m2 * (Ie1**a2 - 3**a2) - (m1 + m2) * ln(Jstar / Jv)
    psiE_eq = 0.5 * (mk - veps) * I4 - 0.5 * mk * I5
    psiE_Neq = 0.5 * (nk - eps) * Ie4 - 0.5 * nk * Ie5
    W_hat = psi_Eq + p * (J - Jstar) + psi_Neq + psiE_eq + psiE_Neq
    return W_hat
\end{lstlisting}
\begin{lstlisting}[language=Python, caption=The RHS of the evolution equation for the internal variable \eqref{Evol-Eq-Cv3}]
from ufl import Identity, det, grad, tr, inner, ssqrt, ln

def evolEqn(C, Cv, params):
    (mu1, mu2, alph1, alph2, m1, m2, a1, a2, K1, K2, bta1, bta2, eta0, etaInf, mk, nk, eps, veps, zeta, lam1) = params
    Cv_inv = inv(Cv)
    Iv1 = tr(Cv)
    Ie1 = inner(C, Cv_inv)
    Ie2 = 0.5 * (Ie1**2 - inner(Cv_inv * C, C * Cv_inv))
    c_neq = m1 * (Ie1 / 3.0) ** (a1 - 1.0) + m2 * (Ie1 / 3.0) ** (a2 - 1)
    J2Neq = (Ie1**2 / 3 - Ie2) * c_neq**2
    etaK = etaInf + (eta0 - etaInf + K1 * (Iv1**bta1 - 3.0**bta1)) / (1 + (K2 * J2Neq) ** bta2)
    G = (c_neq / etaK) * (C - Ie1 * Cv / 3)
    return G
\end{lstlisting}

% \begin{lstlisting}[caption={evolEqn and k_tE Functions}, label={lst:evolEqn_k_tE}]

% def k_t(dt, u, un, Cvn, params, dCv):
%     F = Identity(len(u)) + grad(u)
%     C = F.T * F
%     ufl_form_Cv = (
%         inner(
%             Cv - Cvn - dt * evolEqn(C, Cv, params),
%             dCv,
%         )
%         * dx
%     )

%     return ufl_form_Cv
% \end{lstlisting}

% \{No Ev evolution}
% \paragraph{RK5}
% \paragraph{Backward Euler}
% % \paragraph{Adams??}
% a) Compare results in plots, b) comment on the time/time steps, c) Comment on the inefficiency\\

% All of these cases and results indicate that RK5 is not suited for the time discretization of this problem due to high contrast in the mechanical and dielectric relaxation times of dielectric elastomers. \footnote{it is assumes that the time scale is much larger than the relaxation time of dielectric response}

% \section{Comparison with Experiments: VHB 4910}
% \input{Experiments/comparison_experiments}

\section{Final Comments}\label{sec:Final_Comments}
A stable and robust numerical framework to simulate the nonlinear electromechanical response of dielectric elastomers is presented. A thorough investigation into three time-integration schemes to integrate the evolution equation for the internal variables is presented. Counter to general expectation, and previously reported results, it is found out that low-order implicit schemes are often robust in dealing with the disparate time-scales of electric and mechanical dissipation. Moreover, high-order explicit time-integration schemes often lead to unrealistically small time-increments and hence are not suitable in practice. A parallel numerical framework combining the time-integration schemes together with a conforming finite-element discretization is presented in the open-source framework \texttt{FEniCSx}. The scheme is deployed to simulate the nonlinear electromechanical response of a commercial dielectric elastomer (VHB4910) under combined electromechanical loadings.

While the constitutive model in \cite{ghosh2021two} is general enough to describe the response of a large class of elastomers, there is room for improvement especially in the context of data-driven constitutive modeling. Recently data-driven constitutive modeling, especially for problems involving viscous dissipation or damage (plasticity), has received considerable attention (see, e.g. \cite{tacc2023data,guan2023neural}). Our open-source FE implementation can be tightly integrated with data-driven constitutive models as well as existing Machine Learning models such as Neural Operators (\cite{li2020fourier,lu2019deeponet}) to develop ML surrogates that can learn to simulate the electromechanical response of dielectric elastomers.

On a separate front, the constitutive model presented in this paper can be generalized to other time-dependent behaviors, such as those found in liquid-metal filled elastomers that exhibit interface mechanics (surface tension) at the inclusion-elastomer interface. This can be thought of as a generalization of the work presented in \cite{ghosh2022elastomers}.
\\
\\
\textbf{Acknowledgments}: The authors would like to express their sincere gratitude to \href{https://cbe.ncsu.edu/people/mddickey/}{Prof. Michael Dickey} and \href{https://mechse.illinois.edu/people/profile/martinos}{Prof. Martin Ostoja-Starzewski} for their invaluable inputs in this study. Their expertise and insights were instrumental in shaping some of the topics of this study. B S would also like to thank \href{https://theorg.com/org/palfinger/org-chart/matthias-rambausek}{Matthias Rambausek} for insightful discussions that lead to the implementation in this work.

% and finally the 
% Why needed??
% \begin{enumerate}
%     \item Parallelization
%     \item Can be instantiated as a class for parameter identification??
%     \item for PINN/CANN for material characterization (in companion subsequent papers)
% \end{enumerate}
% Also, CANN? 
% Companion paper to future publications: a) CANN/PINN and b) the non-linear time-dependent behavior of liquid-metal filled elastomers that exhibit interface mechanics (surface tension) at the inclusion-elastomer interface. The latter would be a generalization of the work presented in \cite{ghosh2022elastomers}.

\newpage
\bibliography{reference}
\bibliographystyle{plainnat}
\newpage

\appendix
% \section{Appendices}
% \section{Staggered Solver}\label{Appendix:Staggered_Solver}
%
\tikzstyle{startstop} = [rectangle, rounded corners, minimum width=3cm, minimum height=1cm, text centered, draw=black, fill=red!30]
\tikzstyle{process} = [rectangle, minimum width=3cm, minimum height=1cm, text centered, draw=black, fill=orange!30]
\tikzstyle{decision} = [diamond, aspect=2, text centered, draw=black, fill=green!30]
\tikzstyle{arrow} = [thick,->,>=stealth]

\section{Material Parameters for VHB 4910}\label{Sec:Material_Parameters}

\begin{table}[h!]
\caption{Material parameters for VHB 4910 determined from experiments in \cite{qiang2012experimental} and \cite{hossain2012experimental}
}\centering
\begin{tabular}{ll}
\hline
\hline
$\mu_1=13.54 \,{\rm kPa}$ &  $\mu_2=1.08 \,{\rm kPa}$ \\
$\alpha_1=1.00$ & $\alpha_2=-2.474$   \\
$\nu_1=5.42\,{\rm kPa}$ &  $\nu_2=20.78 \,{\rm kPa}$ \\
$\beta_1=-10$ & $\beta_2=1.948$    \\
\hline
$\eta_0=7014\,{\rm kPa} \cdot {\rm  s}$  &  $\eta_\infty=0.1 \,{\rm kPa} \cdot{\rm  s}$  \\
$\gamma_1=1.852$ &  $\gamma_2=0.26$  \\
$K_1=3507 \,{\rm kPa} \cdot{\rm  s}$  & $K_2=1\,{\rm kPa}^{-2}$ \\
\hline
$\varepsilon=4.48 \varepsilon_0$  &  $m_{\texttt{K}}=3.08 \varepsilon_0$  \\
$\epsilon=-2.68 \varepsilon_0$ &  $n_{\texttt{K}}=-0.2788 \varepsilon_0$  \\
\hline
$\zeta=3.69\varepsilon_0\times 10^{-6}\,{\rm s}$  &  $\,$  \\
\hline
\hline
\end{tabular} \label{Table1}
\end{table}
%

% \section{Abaqus UEL implementation}\label{Sec:Abaqus}
% \input{Abaqus_UEL_Implementation/abaqus}

%
\end{document}